\documentclass[letterpaper]{article} 
\usepackage{aaai2026}  
\usepackage{times}  
\usepackage{helvet}  
\usepackage{courier}  
\usepackage[hyphens]{url}  

\usepackage{graphicx} 
\usepackage{natbib}  
\usepackage{caption} 
\usepackage{algorithm}
\usepackage{algorithmic}

\usepackage{newfloat}
\usepackage{listings}
\DeclareCaptionStyle{ruled}{labelfont=normalfont,labelsep=colon,strut=off} 
\floatstyle{ruled}
\newfloat{listing}{tb}{lst}{}
\floatname{listing}{Listing}
\usepackage{xcolor}
\newcommand{\answerYes}[1]{\textcolor{blue}{#1}} 
\newcommand{\answerNo}[1]{\textcolor{teal}{#1}} 
\newcommand{\answerNA}[1]{\textcolor{gray}{#1}}

\newif\ifreview
\reviewfalse 
\ifreview
  \newcommand{\rev}[1]{\textcolor{blue}{#1}}
\else
  \newcommand{\rev}[1]{#1}
\fi

\title{Don't You Know, Pump it Up! \\Investigating Cryptocurrency Manipulation in Telegram-Driven Activity}
\author {
    Filipe Moura\textsuperscript{\rm 1},
    Giordano Paoletti\textsuperscript{\rm 2},
    Carlos H.G. Ferreira \textsuperscript{\rm 3},
    Jussara M. Almeida\textsuperscript{\rm 1}
    }
\affiliations {
    \textsuperscript{\rm 1} Universidade Federal de Minas Gerais, Belo Horizonte, Brazil\\
    \textsuperscript{\rm 2} Politecnico di Torino, Turin, Italy\\
    \textsuperscript{\rm 3}  Universidade Federal de Ouro Preto, Ouro Preto, Brazil\\
        \{filipe.moura,jussara\}@dcc.ufmg.br, giordano.paoletti@polito.it, chgferreira@ufop.edu.br
}

\usepackage{bibentry}
\usepackage{xspace}
\usepackage{amsmath}
\usepackage{booktabs}
\usepackage{subcaption}
\usepackage{graphicx}
\usepackage[most]{tcolorbox}
\usepackage{amssymb}

\newtcolorbox{msgbox}[2][]{
    colback=gray!3,
    colframe=gray!80!black,
    fonttitle=\bfseries\small,
    title=#2,
    arc=1mm,
    boxrule=0.5pt,
    left=2mm, right=2mm, top=1mm, bottom=1mm,
    before skip=10pt, after skip=10pt,
    #1
}

\newcommand{\pump}{\textit{Pump-and-dump}\xspace}

\makeatletter
\let\savedtitle\@title
\gdef\@title{%
\vspace*{-5em}
  \makebox[\textwidth][c]{%
    \setlength{\fboxsep}{7pt}%
    \fcolorbox{black}{yellow!15}{%
      \parbox{0.93\textwidth}{\normalfont\small\raggedright
    \textit{\textbf{Accepted at ICWSM 2027}. If you cite this paper, please use the
    following reference:} Moura, Filipe; Paoletti, Giordano; Ferreira, Carlos H.~G.;
    and Almeida, Jussara M. 2027. \emph{Don't You Know, Pump It Up! Investigating
    Cryptocurrency Manipulation in Telegram-Driven Activity.} In \emph{Proceedings
    of the 21st International AAAI Conference on Web and Social Media} (To appear).
      }%
    }%
  }%
  \par\vspace{1.2em}%
  \savedtitle
}
\makeatother

\begin{document}

\setlength{\textfloatsep}{3pt}
\setlength{\abovedisplayskip}{3pt}
\setlength{\belowdisplayskip}{3pt}
\setlength{\abovedisplayshortskip}{1pt}   
\setlength{\belowdisplayshortskip}{1pt}

\maketitle


\begin{abstract}

Telegram plays a pivotal role in cryptocurrency communication and has been repeatedly associated with coordinated schemes, such as pump-and-dump manipulation. However, existing studies typically focus on known manipulation chats or a limited set of cryptocurrencies, leaving open the question of how Telegram is leveraged for mass promotional activity (shilling) at scale. 
Moving beyond these limitations, this work analyzes the interplay between information flows and market activity across public Telegram channels. 
To this end, we propose a scalable framework that (i) classifies crypto-related messages using a fine-tuned encoder model to filter semantic noise, (ii) detects anomalous spikes in cryptocurrency mentions via adaptive thresholding, and (iii) validates temporal associations between social bursts and market movements using quasi-experimental econometric methods (RDD and DiD).
We apply this framework to one year of public Telegram data (14,499 channels and over 20 million messages) aligned with transaction data for more than 17,000 cryptocurrencies. Our analysis identifies 47 events consistent with potential pump-and-dump activity and 73 sustained market reactions, showing that manipulative signals are characterized by extreme temporal synchronization and precede price movements by seconds. Notably, psycholinguistic analysis reveals that pump-and-dump messages are linguistically indistinguishable from organic discussions, highlighting the limits of text-based detection alone. Finally, we estimate the cumulative financial volume of detected pump-and-dump events to exceed \$200 million and release a public cryptocurrency dictionary and a fine-tuned classifier to support future research.\footnote{
\textbf{References:}\\
\textbf{Code} \url{https://github.com/Filipey/investigating-cryptocurrency-manipulation-icwsm2027}\\
\textbf{Dataset} \url{https://huggingface.co/datasets/filipeasm18/crytpo-related-labeling}\\
\textbf{Dictionary} \url{https://doi.org/10.5281/zenodo.22116582}\\
\textbf{Model} \url{https://huggingface.co/filipeasm18/roberta-binary-crypto-related-classifier}
}

\end{abstract}

\section{Introduction}\label{sec:intro}

Telegram has become a central communication infrastructure for cryptocurrency communities, enabling rapid information diffusion through large public channels (no membership limits) and other high-reach spaces. At the same time, its scale and limited oversight create a fertile ground for institutionally illegal or ethically questionable activities  \cite{LaMorgia:2021, Roy:2024} and have repeatedly been associated with coordinated market manipulation, most notably \pump schemes, in a market context characterized by high volatility and decentralization~\cite{Mezquita:2022}. While such schemes have been documented \cite{Hu:2023, Clough:2023}, we still lack a clear ecosystem-level understanding of how broadly public Telegram channels are involved in their coordination beyond a small set of known manipulation chats and highly visible assets.

Cryptocurrency markets are particularly sensitive to information flows, as prices react rapidly to speculative signals and social attention~\cite{Krishnan:2024}. Social media platforms, and Telegram in particular, play a key role in disseminating such information, enabling the coordination of large numbers of participants around specific assets~\cite{Mirtaheri:2021,Amirzadeh:2023}. Within this environment, coordinated schemes such as \pump operations -- where collective buying is used to artificially inflate prices before a rapid sell-off-- have emerged as a recurrent form of market manipulation, and their orchestration on Telegram and other platforms has been widely documented~\cite{LaMorgia:2023}.

Prior work has extensively studied \pump schemes by analyzing transaction data, sometimes in combination with social media signals, to characterize their dynamics \cite{Dhawan:2022,Nizzoli:2020}, detect suspicious events, or predict their success~\cite{Amirzadeh:2023,Krishnan:2024}. However, these studies typically focus on a limited set of predefined cryptocurrencies or on Telegram chats already known to orchestrate manipulation. As a result, we still lack an ecosystem-level view of how coordinated \pump activity manifests across Telegram more broadly, including beyond explicitly crypto-focused channels.

In this study, we analyze public Telegram communication alongside cryptocurrency market data to study how coordinated \pump activity emerges on the platform. Rather than focusing on predefined assets or known manipulation chats, our approach examines the relationship between information flows and financial transactions across Telegram channels, enabling the identification of suspicious coordination without prior assumptions about the coins or channels involved. Moreover, since crypto-related discussions frequently occur in non-crypto chats~\cite{perlo2025topic,Paoletti:2025}, we extend our analysis to Telegram channels spanning diverse topical domains. 

\rev{Drawing on information cascade theory and market microstructure literature, we hypothesize that successful \pump schemes may operate in two distinct phases: a private coordination phase (in closed groups) and a public amplification phase (in open channels) to generate sufficient exit liquidity. Regardless of how a \pump scheme is initially orchestrated, this implies that public Telegram channels carry a systematic, temporally compressed signal that is observable even without access to private orchestration chats.} \rev{In this study, public Telegram channels are treated as observable traces of promotional amplification and shilling activity, rather than as direct evidence of the private coordination mechanisms that may precede them}.
Our goal is to characterize Telegram’s role in potential \pump schemes by presenting evidence of temporal associations between crypto-related message volumes and market prices. To this end, we structure our study around three research questions:

 \noindent\textbf{RQ1:} How can potential \pump attempts be surfaced from large-scale Telegram activity and assessed through subsequent market behavior?


\noindent\textbf{RQ2:} When do anomalous increases in cryptocurrency mentions on Telegram coincide with short-term market events consistent with \pump activity?

\noindent\textbf{RQ3:} How do communication patterns of Telegram differ between \pump events and periods of increased cryptocurrency discussion driven by organic market movements (transaction volume and price shifts)?
To address these questions, we propose a three-step analytical framework designed to uncover and validate patterns consistent with \pump activity on Telegram. First, we build a crypto-related classifier to accurately filter messages mentioning cryptocurrencies, addressing naming ambiguities among viral \textit{altcoins} (e.g., \texttt{\$Trump}, \texttt{\$Jesus}) whose polysemic names complicate keyword-based detection~\cite{Anderson:2024}. Second, we apply the \textit{Cell-Average Constant False Alarm Rate} (CA-CFAR) method to detect anomalous increases in daily cryptocurrency mentions, identifying candidate days for subsequent minute-level analysis~\cite{Richards:2010,Belluta:2023}. Third, we employ a Regression Discontinuity Design (RDD) combined with Difference-in-Differences (DiD) to examine temporal associations between these social bursts and market behavior~\cite{Lee:2010}, using stablecoins and Bitcoin as a control group to filter systemic noise and distinguish \pump attempts from other forms of market activity.
We then characterize the market impact and communication patterns of the validated events to assess the potential role of Telegram in market dynamics and how communication during these events differs from legitimate discussions.

In sum, our main contributions are fourfold:
(1) a scalable analytical framework that enables the study of \pump coordination on Telegram beyond predefined assets or known manipulation chats;
(2) a large-scale empirical analysis of public Telegram activity and cryptocurrency market data, providing evidence of when anomalous social attention aligns with market behavior consistent with \pump activity;
(3) the release of a public dictionary containing over 18,704 cryptocurrency-related terms (17,012 coins) to support future research in this area; and
(4) a robust crypto-related text classifier and a labeled dataset of 2,923 Telegram messages for model training and evaluation.

\section{Background and Related Work}\label{sec:background}
In this section, we review the main forms of cryptocurrency market manipulation and summarize prior studies on \pump schemes.
\subsection{Cryptocurrency Manipulation}

Cryptocurrency markets are subject to various types of manipulations, including wash trading \cite{cong2023crypto, Cui:2023}, Ponzi schemes \cite{Bartoletti:2020, Liang:2025}, opportunistic blockchain transaction reordering
\cite{Alipanahloo:2024}, and coordinated NFT market manipulation \cite{Sifat:2024, Niu:2024}. 
Among these, \pump schemes remain one of the most prevalent forms of coordinated manipulation  \cite{Victor:2019, Nghiem:2021,  Dhawan:2022}.

Briefly, the classic \pump scheme consists of three main steps. First, a target cryptocurrency is selected, typically characterized by low market capitalization and high volatility~\cite{Xu:2019}. Second, participants coordinate high-volume purchases of the target asset (the \textit{pump-coin}) to artificially inflate its price. Finally, once the desired price level is reached, early participants initiate the \textit{dump} phase by selling their holdings at a profit, causing a rapid price collapse that leaves later investors with significant losses~\cite{Xu:2019}.

\pump schemes rely on a large number of participants to generate sufficient trading volume, thus attracting new investors and driving the price of the target cryptocurrency. Prior work has documented the orchestration of such schemes on major social media platforms \cite{Nghiem:2021, LaMorgia:2023}, notably Telegram \cite{Nizzoli:2020, Clough:2023,  Chen:2023}, where group/channel administrators orchestrate synchronized message bursts to attract buyers and profit as early sellers \cite{Xu:2019}.

Grounded in market microstructure and behavioral finance theories of manipulation, we hypothesize that while the initial orchestration of these schemes may occur in private chats, their successful execution inherently requires massive public exposure to generate exit liquidity and attract external investors with \textit{Fear of Missing Out} (FOMO). Therefore, monitoring public channels provides a distinct, highly synchronized temporal footprint of the operational phase of market manipulation, especially on \pump activities.

\subsection{Prior Studies on \pump Schemes}



Some prior studies of \pump schemes aimed at characterizing the manipulation ecosystem. For example, prior work suggested that Twitter can act as a gateway for the coordination of \pump and Ponzi schemes \cite{Nizzoli:2020}, and can significantly influence cryptocurrency price appreciation \cite{Amirzadeh:2023}.
Others focused on pump events orchestrated on Telegram, identifying 
the targeted cryptocurrencies and reporting long-term value drops \cite{Clough:2023, Chen:2023}. Beyond classic pump groups, manipulative behavior has also been observed in Initial Exchange Offerings (IEOs), a fundraising mechanism conducted on exchanges, where bot-driven Twitter hype correlates with short-term volatility and pump-like dynamics \cite{Tian:2024}.


Another body of research aimed at detecting \pump events by, for instance, leveraging Telegram messages to identify affected coins on major exchanges using transaction-based classifiers \cite{Victor:2019}.  Multi-platform approaches, involving Telegram, Twitter and Discord, have also been proposed \cite{Mirtaheri:2021, LaMorgia:2023}. 
Additional work showed that tailored oversampling and learning strategies can detect manipulation signals prior to the pump peak \cite{Fantazzini:2023}. Recently, temporal graph neural networks with contrastive learning was applied on transaction data for real-time  \pump detection\cite{Wu:2025}.

Another line of research focused on prediction models, by using Random Forests  with transaction data as input features to predict \pump success in real time \cite{LaMorgia:2020}, or leveraging neural networks to estimate  the maximum value reached by a pump \cite{Nghiem:2021}. Additional approaches combined web data (e.g., social media signals and Google searches) with transaction data to predict price fluctuations and trading behavior \cite{Boukhers:2023, Krishnan:2024}. 

Our work differs from prior studies by addressing open questions about how \pump activity manifests on Telegram at scale.
First, our approach enables the identification of cryptocurrencies potentially targeted by \pump manipulation on Telegram without preselecting specific coins, exchanges, or channels as orchestrators, unlike \cite{Victor:2019,Mirtaheri:2021,Clough:2023}. By not restricting the analysis to cryptocurrency-focused channels, and acknowledging that relevant messages may appear outside explicitly crypto-related chats~\cite{Paoletti:2025,perlo2025topic}, we capture \pump-related activity across diverse Telegram channel types. 
Second, we study how communication patterns on Telegram differ during \pump events compared to periods of increased discussion driven by organic market movements.




\section{Methodology}\label{sec:meth}

\begin{figure*}
    \centering
    \includegraphics[width=0.8\linewidth]{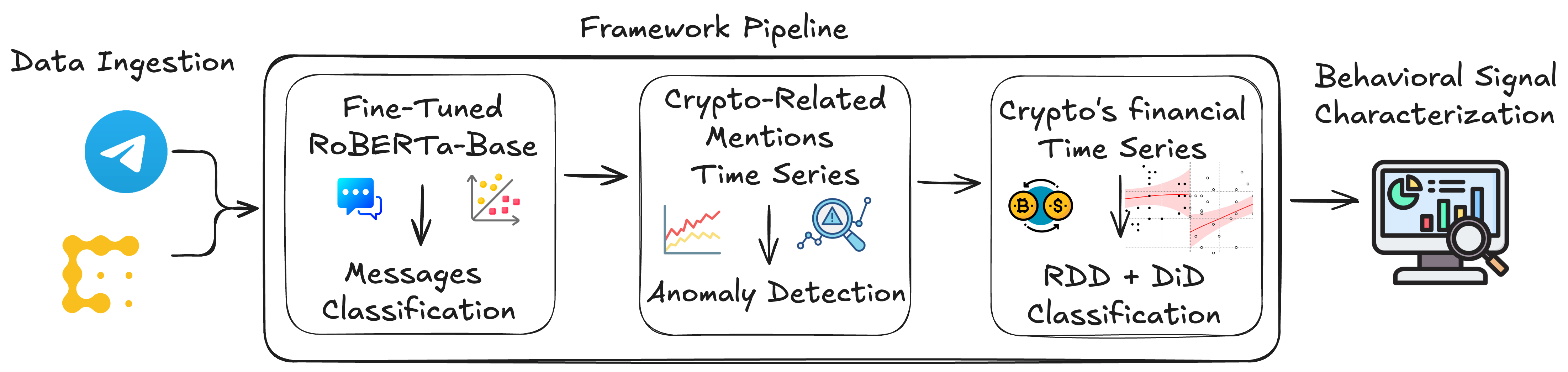}

    \caption{Pipeline for revealing possible \pump occurrences driven by Telegram activities. }
    \label{fig:methodology_pipeline}
    
\end{figure*}

To investigate the interplay between online discussions and cryptocurrency markets, we introduce an analytical framework tailored to Telegram data. 
Our goal is the characterization of patterns consistent with \pump manipulation emerging from Telegram activity and the understanding of how such activity intertwines with cryptocurrency market dynamics. 
Specifically, we investigate whether Telegram discussions precede, coincide with, or follow market fluctuations, highlighting temporal links between social and market dynamics.
To achieve this, we design a three-step pipeline that progressively refines the analysis, from raw message streams to fine-grained temporal patterns, ensuring both scalability and analytical precision.
The pipeline, illustrated in Fig.~\ref{fig:methodology_pipeline},  consists in: 
\textbf{(1)} building a classifier to accurately filter crypto-related content; \textbf{(2)} detecting anomalous spikes in the daily volume of Telegram messages mentioning cryptocurrencies; \textbf{(3)} uncovering \textit{Telegram–market} temporal association to verify correlation links and assess whether suspicious Telegram bursts acts as a signal for trading orders or are an organic behavior tailored by a market trend. Such temporal associations help characterize Telegram’s role in market dynamics, revealing when activity patterns are more consistent with potential manipulation signals or reactive market responses. 

The framework takes two primary data sources as input. First, for Telegram, \rev{we leverage the TGDataset~\cite{LaMorgia:2025}, sampling its final year (July 2021 -- July 2022): Over 20 million messages across 14,499 public channels}. Second, we utilize a dictionary of cryptocurrencies, each with a corresponding time series of transaction volumes and prices from one or more exchanges, temporally aligned with the Telegram data. Unlike prior work, this cryptocurrency dictionary is intentionally broad and not restricted to assets previously linked to \pump events. This lack of restriction allows for the unbiased identification of coins exhibiting patterns consistent with potential manipulation. We explain each step of our framework in detail next.

\subsection{Cryptocurrency Content-Related Classifier}
Before identifying days with an anomalous volume of mentions for a cryptocurrency, it is necessary to validate the message's content due to possible polysemy caused by the coin's name. For instance, coins such as \texttt{Jesus}, \texttt{Skate}, and \texttt{Trump} introduce semantic ambiguity when relying solely on keyword-based searches, potentially yielding false positives from discussions unrelated to cryptocurrencies. A semantic classifier is therefore not merely a preprocessing convenience but a necessary design choice. To account for message context, we employ an experimental setup to evaluate the usage of Large and Small Language Models (SLMs and LLMs) to classify captured messages as \textit{crypto-related} or \textit{not}, a new challenge that arises from the abundant \pump occurrences search across a broad Telegram snapshot.

\subsubsection{Data Annotation}
To establish a reliable ground truth for identifying cryptocurrency-related discourse, we curated a dataset of 2,923 messages. This corpus was strategically sampled to capture the diverse linguistic nuances of the ecosystem: \begin{itemize} \item \textbf{High-Risk Domain Specific (463 messages):} Targeted samples from assets historically prone to manipulation (e.g., memecoins and confirmed Pump-and-Dump cases such as \texttt{\$HAWK}\footnote{https://www.bbc.com/news/articles/c89xvjkzzyvo} and \texttt{\$LIBRA}\footnote{https://www.bbc.com/news/articles/cp9x9j89evxo}). \item \textbf{Random Lexical Sampling (1,960 messages):} Obtained using a comprehensive 18,704-term cryptocurrency dictionary to capture general organic discourse. \item \textbf{Manipulative Context (500 messages):} Sourced from known predatory groups during confirmed manipulation windows as defined by \citet{LaMorgia:2020}. \end{itemize}

The annotation process was conducted by three Master's students in Computer Science. Given the technical nature of the task and the absence of external domain specialists, annotators followed a rigorous binary classification protocol: \textit{``A message is crypto-related if it mentions Cryptocurrencies (Bitcoin, Ethereum, altcoins, tokens), NFTs, blockchain, Web3, DeFi, exchanges, wallets, mining, staking, airdrops, crypto trading, investing, tokens, presales, jargons or memecoins. Even a brief or indirect mention is sufficient''}. To ensure the reliability of the resulting ground truth, we calculated the Fleiss' Kappa and Krippendorff's Alpha score to quantify inter-annotator agreement before adopting the majority-vote labels for subsequent model evaluation \cite{McHugh:2012}.





\subsubsection{Experimental Configuration}
Our experimental design aims to identify the optimal balance between computational efficiency and classification effectiveness. We selected a heterogeneous set of models to represent the current state-of-the-art in both Large Language Models (LLMs) and encoder-based architectures: \texttt{Llama 3.1 8b-Instruct}, \texttt{Gemma 7b v1}, \texttt{Qwen2.5 7b}, and \texttt{RoBERTa-Base} \cite{Liu:2019, Dubey:2024, Team:2024, Bai:2025}. The choice of these specific models is justified by their architectural diversity and performance benchmarks. \texttt{Llama 3.1}, \texttt{Gemma}, and \texttt{Qwen} represent top-tier decoder-only models with strong instruction-following capabilities, optimized for varying parameter scales that remain feasible for institutional hardware. In contrast, \texttt{RoBERTa-Base} was selected as a discriminative baseline; its bidirectional encoder architecture is traditionally more efficient for sequence classification tasks with lower computational overhead compared to generative LLMs. Each model was evaluated across four distinct learning paradigms: \begin{enumerate} \item \textbf{Zero-Shot Classification: (ZSC)} The model classifies messages based solely on its pre-trained knowledge without any specific instructions. \item \textbf{Zero-Shot In-Context Learning (ZS-ICL):} The classification is guided by the inclusion of the official annotation guidelines within the prompt. \item \textbf{Few-Shot In-Context Learning (FS-ICL):} The prompt includes five randomly selected exemplars (3 positive, 2 negative) to provide the model with a linguistic pattern for comparison, following best practices in recent literature \cite{Cunha:2025}. \item \textbf{Fine-Tuning (FT):} The models undergo supervised weight adjustment on a dedicated training split of the annotated dataset, representing the upper bound of domain-specific adaptation. \end{enumerate}

To ensure the robustness and generalizability of the Fine-Tuning results, we implemented a 5-fold Cross-Validation scheme (80/20 split), mitigating potential biases in the training distribution. For ICL methods, we maintained a unified prompt template and employed deterministic decoding (e.g., temperature set to 0) to ensure fair and reproducible comparisons between LLMs; the full prompt architecture is detailed in Appendix A\ref{sec:app-llm}. Beyond standard effectiveness metrics, specifically, Recall, Precision, F1-Score, and Accuracy \cite{Cunha:2025},  we conducted an efficiency analysis by measuring mean inference time. This dual-metric approach allows for identifying the optimal model that achieves high classification performance without prohibitive computational costs for possible real-time monitoring of large-scale Telegram data.

\rev{To further ensure the temporal robustness and generalization of our annotations, we conducted an additional validation experiment. We extracted a temporally stratified sample of 520 messages spanning 13 months (July 2021 to July 2022), balanced by the model's predictions (predicted positive vs. negative). By blinding human annotators to the model's predictions, we evaluated the False Positive (FP) and False Negative (FN) rates, confirming that the classifier maintains stable accuracy across time without introducing systematic bias. More details are available in Appendix A \ref{sec:app-llm}.}

\subsection{Detection of Anomalous Spikes in Telegram}
\label{subsec:mention_anomaly_detection}
After training the crypto-related classifier, our framework proceeds by identifying days exhibiting an {\it anomalous} volume of messages mentioning a cryptocurrency.
Specifically, Telegram messages are first matched against a comprehensive cryptocurrency dictionary to collect candidate mentions, and are then filtered using the trained classifier to remove false positives due to polysemy. Only after this semantic filtering step do we construct time series of daily mention counts.
For each cryptocurrency $m$ in the dictionary, we count the daily number of messages containing its official ticker (i.e., $\$m$) and its primary variants ("$m$", "$m$ coin", and "$m$ token"). This count is computed across all channels, regardless of their specific topic, as prior work has shown that cryptocurrency-related information often transcends crypto-specific channels~\cite{Paoletti:2025}.

Given the significant heterogeneity in the popularity of cryptocurrencies, an effective anomaly detection method must adapt to the specific context of each coin rather than relying on a single global threshold. Therefore, in this work, we adopt a time-series anomaly detection method known as \textit{Cell-Average Constant False Alarm Rate} (CA-CFAR) \cite{Richards:2010}. This is one of several implementations of the CFAR (\textit{Constant False Alarm Rate}) algorithm, which has previously been employed for detecting activity \textit{bursts} in social media \cite{Belluta:2023}.

CFAR adjusts a detection threshold adaptively to maintain a constant rate of false alarms. Specifically, for a given time series, CFAR uses a sliding window to evaluate each data point, referred to as the Cell Under Test (CUT). It then estimates a threshold $T$ based on the signal magnitude in the cells neighboring the CUT\footnote{https://www.mathworks.com/help/phased/ug/constant-false-alarm-rate-cfar-detection.html}. The daily threshold is defined as $T = \alpha P_n$, where $P_n$ represents an estimate of the background noise derived from neighboring cells, and $\alpha$ is a scaling factor controlling the sensitivity of the detector.

The background estimate $P_n$ is computed from cells adjacent to the CUT on both sides of the time series. A sliding window centered on the CUT is applied, parameterized by $N$ \textit{training} cells and $M$ \textit{guard} cells on each side. The \textit{guard} cells, located immediately adjacent to the CUT, are excluded from the noise calculation to prevent the target signal from contaminating the background estimate \cite{Richards:2010}. CA-CFAR estimates the background level separately for the left and right sides of the CUT and then averages the two estimates.

Formally, let $x_t$ be the value of the time series on day $t$, and let $d$ denote the direction relative to the CUT (\textit{left}: preceding days; \textit{right}: subsequent days). Let $N_d$ and $M_d$ denote the number of training and guard cells in direction $d$, respectively. For a CUT at day $t$, the directional background estimate is defined as:
\[
P_d(t) =
\frac{1}{N_d}
\sum_{k=M_d+1}^{M_d+N_d} x_{t+s_d k},
\qquad
d \in \{\text{left},\text{right}\},
\]
where $s_{\text{left}}=-1$ and $s_{\text{right}}=+1$. In our symmetric setting, $N_{\text{left}}=N_{\text{right}}=N$ and $M_{\text{left}}=M_{\text{right}}=M$. The final background-noise estimate is then computed as $P_n(t) = \tfrac{1}{2}\left(P_{\text{left}}(t) + P_{\text{right}}(t)\right)$.

The CA-CFAR algorithm computes the scaling factor $\alpha$ from a given Probability of False Alarm ($P_f$), which is interpreted as the expected fraction of data points classified as anomalies. Let $N_{\mathrm{tot}} = N_{\text{left}} + N_{\text{right}}$ denote the total number of training cells used to estimate the background level. The scaling factor is computed as $\alpha = N_{\mathrm{tot}}\bigl(P_f^{-1/N_{\mathrm{tot}}} - 1\bigr)$.
With the values of $P_n(t)$ and $\alpha$, the threshold is computed as $T(t)=\alpha P_n(t)$. For each cryptocurrency time series, we compute $\alpha$, $P_n(t)$, and $T(t)$ on a daily basis to identify anomalous days, i.e., days in which the message volume exceeds the adaptive threshold $T(t)$.

Considering the context of market manipulation, where target coins are often less popular and \pump events are typically short-lived, frequently lasting only minutes or hours \cite{Xu:2019, LaMorgia:2020, Nghiem:2021}, we set the following parameters: $N=5$ training cells on each side of the CUT, $M=1$ guard cell on each side to mitigate noise from post-event discussions \cite{LaMorgia:2023, Clough:2023, Mirtaheri:2021}, and $P_f=0.05$. This represents a conservative approach intended to detect only the most significant deviations from the recent baseline. Higher $P_f$ values increase the number of detections but also elevate the risk of false positives \cite{Richards:2010}. Sensitivity checks around the definition of training and guard cells are explored in Appendix~B \ref{sec:app-cfar}.

\subsection{Telegram-Market Temporal Association}
To rigorously assess the temporal association between social-media-driven information shocks and cryptocurrency market microstructures, we employ a multi-layered econometric strategy. This study goes beyond simple correlation by integrating quasi-experimental designs, specifically, Regression Discontinuity Design (RDD) and Difference-in-Differences (DiD), to control for unobserved market noise and systemic volatility.

\subsection{Data Acquisition and Preprocessing}
To investigate temporal associations, we retrieve high-frequency market data (Open, High, Low, Close, and Volume) for all cryptocurrencies mentioned in the identified Telegram bursts. Data is gathered from the \texttt{CoinDesk Data API} at a 1-minute resolution\footnote{\url{https://developers.coindesk.com}}. This granular scale is essential to capture the rapid onset of \pump events, which often manifest and peak within minutes rather than hours \cite{Xu:2019}. Given the heterogeneous nature of cryptocurrencies, raw prices and volumes are not directly comparable. We therefore transform the raw data into two statistical metrics \cite{Van:2022}:

\begin{enumerate}
    \item \textbf{Accumulated Logarithmic Returns ($R_{\text{acc}}$)}: We convert closing prices into logarithmic returns ($r_t = \ln(P_t/P_{t-1})$), which, unlike simple returns, are time-additive, a property essential for computing cumulative returns consistently across our observation window. The cumulative sum is calculated from $T_{-30}$, centering the series at zero at the start of the analysis to standardize the baseline across all events.
    
    \item \textbf{z-Scored Volume ($z_{\text{vol}}$)}: Trading volume is normalized into z-scores ($z_{\text{vol}} = (V_t - \mu_V) / \sigma_V$) based on the local mean and variance within the analysis window. This metric identifies volume spikes as statistical outliers relative to each asset's own baseline, enabling cross-asset comparison regardless of underlying liquidity, an approach more robust than percentage changes, which can be excessively volatile for low-liquidity assets.
\end{enumerate}

\subsection{Causality Methods}
We adopt the Sharp RDD framework \cite{Lee:2010} to characterize the discontinuity in market behavior at the onset of a Telegram social burst. 
In our setting, once a day is identified as anomalous, the cutoff point ($T_0$) is defined as the timestamp of the first message mentioning the cryptocurrency on that day. Market dynamics are then analyzed at minute-level resolution within a 3-hour window spanning $1.5$ hours before and after $T_0$. 
The RDD exploits the discontinuity at this threshold, assuming that in the absence of the ``treatment'' (the social burst), the market variables would follow a continuous trend. The model is specified as: $Y_t = \beta_0 + \beta_1 D_t + \beta_2(t - T_0) + \beta_3D_t(t - T_0) + \epsilon_t$,
where $D_t$ is a dummy variable indicating the post-burst period ($t \geq T_0$). The coefficient $\beta_1$ represents the instantaneous Jump in price or volume, while $\beta_3$ captures the change in the slope, which we use to quantify the subsequent Reversal Ratio and Volume Decay. To ensure the robustness of our estimates, we performed a sensitivity analysis across a range of bandwidths: $\{10,15,30,45,60,90,120\}$ minutes. We observed that narrow windows (under 30 minutes) often failed to yield statistically significant results due to high-frequency noise and insufficient data points to establish a stable pre-treatment trend. Conversely, while larger windows (above 90 minutes) maintained detection stability, they increased the risk of contamination from exogenous market volatility. Consequently, the 1.5-hour bandwidth was selected as the optimal trade-off, providing consistent significance levels and a reliable baseline for characterizing the immediate impact and subsequent decay of the suspected manipulation. \rev{To rigorously account for potential ``pre-heating'' or early private coordination leaking into the public domain (i.e., that $T_0$ may follow earlier coordination in private or smaller channels not captured in the data), we implemented a placebo test within our RDD framework. We shifted the analysis window to a placebo threshold 3 hours before the actual event ($T_0$-3h), ensuring no temporal overlap with the main RDD window ($\pm1.5$h around $T_0$), which could otherwise introduce cross-contamination bias in the estimates. Any cases exhibiting a statistically significant upward trend before the main burst were systematically filtered out. This ensures our detected $T_0$ captures a genuine, sudden anomaly rather than the continuation of prior momentum}.

A significant challenge to RDD in financial markets is the presence of simultaneous external shocks (e.g., a sudden Bitcoin price swing occurring exactly at $T_0$). To mitigate this, we implement a Difference-in-Differences (DiD) framework using a placebo control group. We synchronize the timestamps of identified anomalies with the market data of major stablecoins (USDT, USDC, DAI) and Bitcoin, as it is the most popular cryptocurrency in terms of market capitalization (\$USD 1.2 trillion), which can guide systematic fluctuations. As the stable assets are pegged to the USD, they are theoretically immune to idiosyncratic \pump coordination, while Bitcoin is effectively unaffected due to its size and liquidity. By comparing the treated altcoin with the control group, we \textit{difference out} systemic noise:
$\delta \text{DiD}
= (Y_{\text{Target}, \text{Post}} - Y_{\text{Target}, \text{Pre}})
- (Y_{\text{Control}, \text{Post}} - Y_{\text{Control}, \text{Pre}})$
A statistically significant $\delta DiD$ that mirrors the RDD jump provides robust evidence that the market anomaly was idiosyncratic to the Telegram activity rather than a reflection of global market trends. The validity of this combined approach relies on three key assumptions \cite{Trochim:1990}: (i) continuity of variables around the cutoff in the absence of the shock, (ii) absence of cutoff manipulation, and (iii) parallel trends between the target coin and the stablecoin/Bitcoin control before $T_0$. 

While alternative approaches, including Interrupted Time Series (ITS) and 
Convergent Cross Mapping (CCM), were also evaluated, the combination of RDD 
and DiD proved the most robust and scalable framework for our setting 
(see Appendix C~\ref{sec:app-causality} for a detailed comparison). Despite 
these controls, we interpret the observed discontinuities as strong 
correlational evidence of social-media-driven market manipulation rather 
than absolute causal proof.

To provide a granular distinction between market phenomena, we classify each event into one of three categories: Insignificant, Sustained Market Reaction, or \pump. This classification is governed by two key behavioral metrics derived from the post-event window: the Reversal Ratio (RR) and the Volume Decay (VD). The categorization criteria are structured as follows: (I) Insignificant: Events where the RDD jump magnitude ($\beta_1$) is below 1\%, or the p-value exceeds $0.05$. This filter eliminates minor fluctuations and statistically weak associations, ensuring that only robust market movements are analyzed. (ii) Sustained Market Reaction: Events that exhibit a significant jump ($>1\%$) but maintain the price level. This is characterized by a Reversal Ratio $\leq0.7$ and a Volume Decay $\geq0.3$, suggesting that the information shock led to a new price equilibrium supported by ongoing trading interest. (iii) \pump (P\&D): Events characterized by a significant price spike followed by a rapid collapse. This is identified by a Reversal Ratio $>0.7$ or a Volume Decay $<0.3$.

The selection of these thresholds is rooted in the principles of Market Microstructure and Efficient Market Hypothesis (EMH) adaptations for high-frequency crypto-assets \cite{Malkiel:1989}. In an efficient market, if an announcement carries fundamental value, the price should stabilize at a new plateau (Price Discovery). A reversal exceeding 70\% of the initial gain implies that the movement lacked fundamental support and was driven by transitory, speculative pressure. We adopt 0.7 as a conservative limit: it allows for minor natural corrections (up to 30\%) while strictly flagging events where the majority of the \textit{wealth} created by the spike was decimated shortly after. This aligns with empirical studies on crypto-manipulation showing that \pump schemes typically return to the mean (mean-reversion) within minutes \cite{LaMorgia:2020, Xu:2019}. Volume is a proxy for market conviction. A sustained reaction is typically accompanied by a \textit{volume hysteresis}, where trading remains elevated as the market processes the new information. Conversely, coordinated manipulation relies on a massive, synchronized burst of orders. A decay below 0.3 indicates that post-spike liquidity has evaporated to less than 30\% of its peak intensity. This \textit{liquidity exhaustion} is a hallmark of artificial activity. Once the coordination (or the bot's execution) ends, the absence of organic interest causes the volume to \textit{dry up}, confirming the initial spike was a non-fundamental artifact. Setting a higher RR (e.g., 0.9) would fail to capture partial dumps where the price settles slightly above the start. At the same time, a lower VD (e.g., 0.1) would be too restrictive, ignoring cases where residual panic-selling maintains some volume. Thus, the 0.7/0.3 threshold pair serves as a robust filter to isolate coordinated anomalies from organic price discovery. \rev{Importantly, these threshold values are heavily theory-driven, reflecting the characteristic behavior of \pump schemes established in the literature. To empirically validate these parameters, we conducted a sensitivity analysis across different threshold configurations (detailed in Appendix C). Overall, event categorization into \pump and SMR remains stable across configurations, with only a small set of borderline cases affected by parameter variation. A targeted manual inspection of these ambiguous cases confirms that the 70/30 configuration minimizes misclassifications, supporting the robustness and practical reliability of our automated separation.}

\section{Results}\label{sec:results}
This section presents our findings, structured to address: identifying suspected pump occurrences (\textbf{RQ1}), establishing temporal precedence between social signals and market shifts (\textbf{RQ2}), and characterizing the information ecosystem involved (\textbf{RQ3}).

\subsection{Crypto-Related Classifier}
The inter-annotator agreement was fair in overral (Fleiss' Kappa/Krippendorff's Alpha =  $0.88$/$0.88$), with stratified results showing higher consensus in \textit{Random Lexical Sampling} ($0.69$/$0.69$) compared to the more ambiguous \textit{High-Risk} ($0.50$/$0.49$) and \textit{Manipulative} ($0.40$/$0.40$) contexts, which underscores the linguistic complexity of predatory discourse. After annotation, the resulting dataset exhibited a class distribution of 76\% non–crypto-related messages and 24\% crypto-related messages. The selection of the primary classification model was based on a dual-objective optimization: maximizing predictive effectiveness (specifically Recall for the minority crypto-class) and minimizing computational overhead for longitudinal large-scale inference. Tables \ref{tab:metrics_c0}, \ref{tab:metrics_c1}, and \ref{tab:time_comparison_update} provide a comprehensive overview of these dimensions. To ensure the reliability of our comparisons, we applied a two-tailed t-test with Bonferroni correction ($\alpha = 0.05$) to all metrics.

Our results reveal a stark performance bifurcation between learning paradigms. Supervised Fine-Tuning (FT) consistently outperformed all In-Context Learning (ICL) and Zero-Shot (ZS) configurations by an average margin of 40 percentage points across F1-scores. Within the FT paradigm, the models achieved near-perfect performance for non-crypto messages ($C_0$), with F1-scores exceeding 97\%. Critically, for the crypto-related class ($C_1$), \texttt{RoBERTa-base} achieved the highest effectiveness, reaching a Recall of 94.0\% and an F1-score of 92.2\%. While \texttt{RoBERTa} numerically surpassed the larger decoder-based models (e.g., \texttt{Llama-3.1-8B} at 91.6\% Recall), the t-test with Bonferroni correction confirmed a statistical tie among all fine-tuned architectures. This finding demonstrates that the inherent bidirectional context of an encoder-only architecture like \texttt{RoBERTa} is as capable as LLMs with $60\times$ more parameters for this specific domain task.

In terms of efficiency, the in-context approaches were evaluated based on inference time during the validation phase. For the fine-tuning experiments, we randomly selected a subset of 618,473 messages, from which approximately 100,000 samples were classified by each model to measure inference time. While the models were statistically equivalent in effectiveness, they were profoundly divergent in computational efficiency.  \texttt{RoBERTa} (FT) proved to be the only viable candidate for real-time monitoring of 17,000 assets. During the validation phase, \texttt{RoBERTa} achieved an average inference latency of 4.78\,ms per message, representing a $13.3\times$ speedup over \texttt{Llama-3.1-8B} (63.63\,ms) and an $18.3\times$ speedup over \texttt{Qwen-2.5-7B} (87.79\,ms); further details are reported in Table~\ref{tab:time_comparison_update} in Appendix A.

The training phase mirrored this efficiency; \texttt{RoBERTa} completed its fine-tuning process in approximately 89 seconds, being $5.6\times$ faster than \texttt{Llama-3.1-8B}. Given that ICL methods-despite avoiding training, incurred prohibitive inference latencies (exceeding 100,000 ms per message in few-shot settings), the fine-tuned \texttt{RoBERTa} model was selected as the optimal backbone for our analysis, combining state-of-the-art predictive power with the throughput necessary for large-scale data processing.
\rev{These results contribute to a growing body of work questioning the assumption that LLMs are the optimal solution across all NLP settings~\cite{bucher2024fine, Cunha:2025}, suggesting that for domain-specific classification under real-world constraints, architectural simplicity may be preferable to scale.}

\begin{table}[t]
\centering
    \footnotesize
\setlength{\tabcolsep}{3.5pt}
\renewcommand{\arraystretch}{1.05}

\begin{tabular}{p{2.62cm}ccc}
\toprule
\textbf{Model/Method} & \textbf{Prec$_0$} & \textbf{Rec$_0$} & \textbf{F1$_0$} \\
\midrule
roberta (FT)        & 98.1(0.3)$\blacktriangle$ & 96.8(0.8)$\bullet$ & 97.5(0.4)$\blacktriangle$ \\
gemma-7b (FT)       & 97.3(1.0)$\bullet$        & 97.3(0.9)$\blacktriangle$ & 97.3(0.4)$\bullet$ \\
llama3.1-8B (FT)    & 97.3(0.8)$\bullet$        & 97.1(1.3)$\bullet$ & 97.2(0.4)$\bullet$ \\
qwen2.5-8B (FT)       & 97.0(0.6)$\bullet$        & 97.1(0.6)$\bullet$ & 97.1(0.2)$\bullet$ \\
\midrule
gemma-7b (ZS)       & 75.88(1.4) & 49.21(1.8) & 59.69(1.6) \\
llama3.1-8B (ZS)    & 74.53(1.5) & 47.95(3.0) & 58.31(2.6) \\
qwen2.5-8B (ZS)       & 76.42(1.9) & 50.38(2.5) & 60.70(2.2) \\
\midrule
gemma-7b (ICL Z)    & 75.91(0.7) & 50.92(2.4) & 60.91(1.6) \\
llama3.1-8B (ICL Z) & 76.75(0.9) & 48.45(2.5) & 59.37(2.1) \\
qwen2.5-8B (ICL Z)    & 76.48(1.0) & 51.69(0.9) & 61.68(0.7) \\
\midrule
gemma-7b (ICL F)    & 75.31(1.9) & 49.35(1.7) & 59.62(1.8) \\
llama3.1-8B (ICL F) & 75.93(1.1) & 48.58(0.8) & 59.24(0.7) \\
qwen2.5-8B (ICL F)    & 75.93(0.7) & 51.10(3.3) & 61.04(2.5) \\
\bottomrule
\end{tabular}

\caption{\footnotesize Non-Crypto-related performance[Avg(Std)]. $\blacktriangle$Best,$\bullet$Tie. }
\label{tab:metrics_c0}
\end{table}

\begin{table}[t]
\centering
\footnotesize
\setlength{\tabcolsep}{3.5pt}
\renewcommand{\arraystretch}{1.05}

\begin{tabular}{p{2.62cm}ccc}
\toprule
\textbf{Model/Method} & \textbf{Prec$_1$} & \textbf{Rec$_1$} & \textbf{F1$_1$} \\
\midrule
roberta (FT)        & 90.5(2.1)$\bullet$ & 94.0(1.0)$\blacktriangle$ & 92.2(1.1)$\blacktriangle$ \\
gemma-7b (FT)       & 91.4(2.4)$\blacktriangle$ & 91.3(3.2)$\bullet$ & 91.3(1.5)$\bullet$ \\
llama3.1-8B (FT)    & 91.0(3.3)$\bullet$ & 91.6(2.6)$\bullet$ & 91.2(1.0)$\bullet$ \\
qwen2.5-8B (FT)       & 90.9(1.5)$\bullet$ & 90.6(1.9)$\bullet$ & 90.7(0.7)$\bullet$ \\
\midrule
gemma-7b (ZS)       & 24.01(1.3) & 50.65(2.8) & 32.58(1.7) \\
llama3.1-8B (ZS)    & 22.76(1.4) & 48.36(3.3) & 30.94(1.8) \\
qwen2.5-8B (ZS)       & 24.54(1.9) & 50.92(4.4) & 33.10(2.6) \\
\midrule
gemma-7b (ICL Z)    & 23.98(0.6) & 48.94(3.7) & 32.17(1.7) \\
llama3.1-8B (ICL Z) & 24.87(0.9) & 53.77(0.9) & 34.00(0.9) \\
qwen2.5-8B (ICL Z)    & 24.60(1.0) & 49.79(3.0) & 32.92(1.6) \\
\midrule
gemma-7b (ICL F)    & 23.43(1.9) & 48.94(4.1) & 31.69(2.6) \\
llama3.1-8B (ICL F) & 24.02(1.1) & 51.35(3.1) & 32.73(1.6) \\
qwen2.5-8B (ICL F)    & 24.10(0.9) & 48.93(2.6) & 32.26(1.0) \\
\bottomrule
\end{tabular}

\caption{\footnotesize Crypto-related performance[Avg(Std)]. $\blacktriangle$Best,$\bullet$Tie.}
\label{tab:metrics_c1}
\end{table}

\rev{The temporal robustness and generalization test indicates stable model performance over time, with a Macro-F1 averaged over 13 months (\textbf{$91.4\%$}) and consistent results across both classes (0 and 1: \textbf{$92.2\%$} / \textbf{$90.5\%$}), with no evidence of temporal drift. We also observe substantial inter-annotator agreement (Fleiss’ Kappa $>>0.6$   across all months), supporting the reliability of these findings (see Appendix A\ref{sec:app-llm}).}

\rev{}
\emph{\textbf{Takeaway} \texttt{RoBERTa} matches or outperforms LLMs in identifying crypto-related messages, while also being computationally scalable for large-scale Telegram analysis.}

\subsection{Detection of Anomalous Market Occurrences}
Initially, our monitoring infrastructure tracked 17,000 unique digital assets across a vast network of Telegram channels. To address \textbf{RQ1} and manage this high-dimensional data, we employed a fine-tuned RoBERTa model, which served as a semantic gatekeeper by isolating strictly crypto-related discourse from general social noise. Once the data were cleaned, the CA-CFAR algorithm identified 9799 bursts of social activity that reached a statistically anomalous threshold, \rev{signaling purely social messaging anomalies.}

\begin{figure}[t]
    \centering

    \begin{subfigure}{0.49\linewidth}
        \centering
        \includegraphics[width=\linewidth]{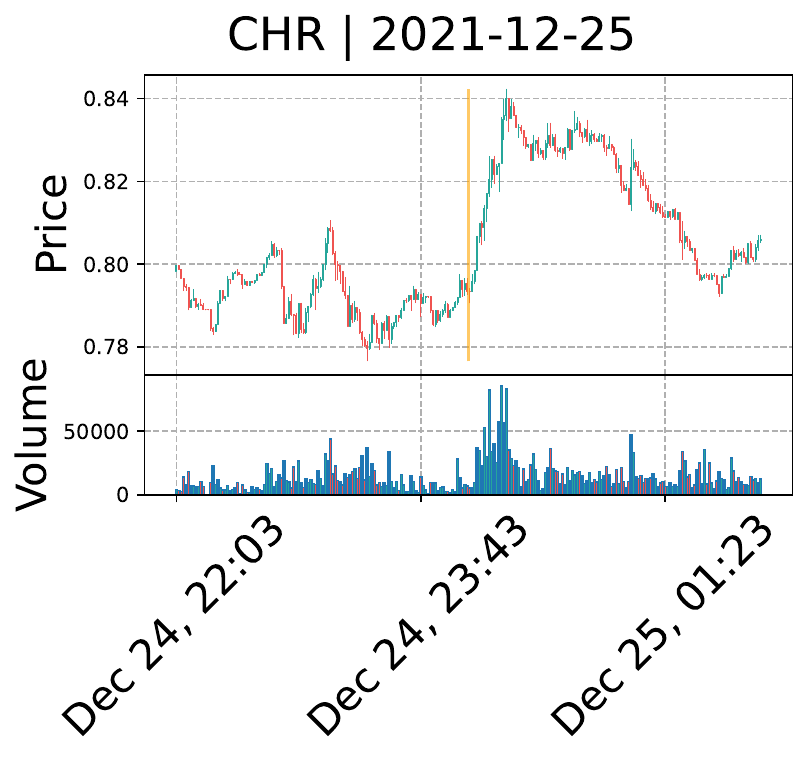}
        \caption{\texttt{\$CHR} \pump.}
        \label{fig:signature-pd}
    \end{subfigure}
    \hfill
    \begin{subfigure}{0.48\linewidth}
        \centering
        \includegraphics[width=\linewidth]{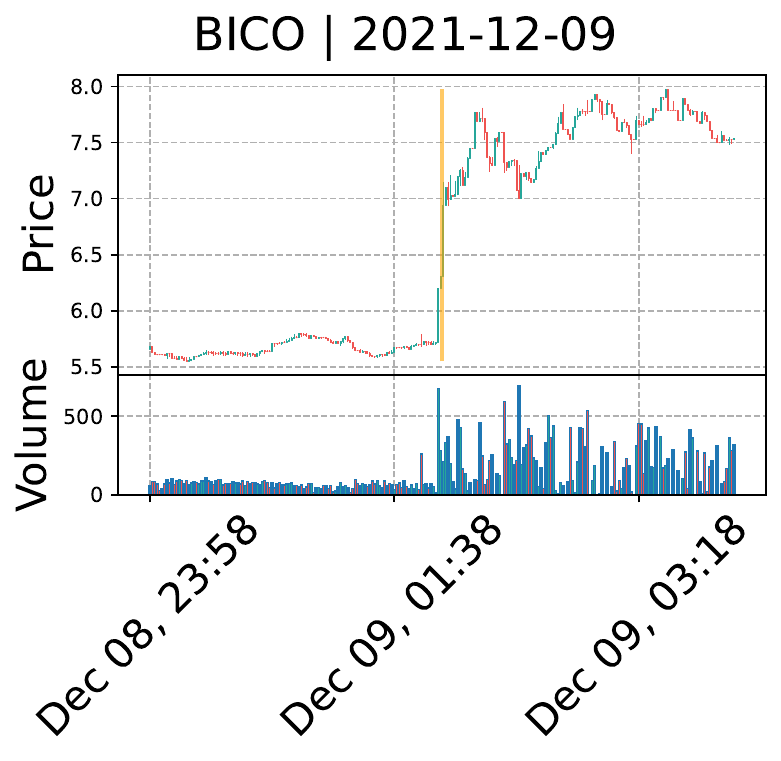}
        \caption{\texttt{\$BICO} market trend.}
        \label{fig:signature-mr}
    \end{subfigure}

    \caption{Signature behavior for P\&D and SMR cases.}
    \label{fig:signature-moves}
\end{figure}

\rev{It is crucial to emphasize that at this stage, no classification of manipulation is made. To properly quantify and classify these events, the social bursts were synchronized with high-frequency market data to test for price discontinuities coincident with the bursts.} 
Using a Regression Discontinuity Design (RDD), we observed 302 immediate price and volume \textit{jumps} following the social signals.
\rev{The  RDD-based placebo test at $T_0-3h$ yielded only 1 statistically significant case, which was filtered out. This near-absence of pre-existing trends supports the validity of our $T_0$ definition as a genuine, sudden anomaly.
}
To ensure these movements were idiosyncratic and not merely reflections of broader market trends, we applied a Difference-in-Differences (DiD) framework using stablecoins and Bitcoin as a control group. This econometric rigor allowed us to discard systemic noise, leaving a core sample of 120 validated high-impact events, comprising 47 \pump (P\&D) cases and 73 Sustained Market Reactions (SMR), separating two kinds of temporal association (\textbf{RQ2}). Figure \ref{fig:signature-pd} exemplifies the trading history for one detected \pump case for the coin \texttt{\$CHR} in 2021. The yellow vertical line represents the detected message burst in Telegram, and the lines indicate positive or negative pricing movement based on the last observation (1-minute interval). The chart's bottom represents the trading volume and type, where blue lines represent buy orders and red lines represent sell orders. This \texttt{\$CHR} case is a classic \pump occurrence, where the target coin's final price gets marginally lower or even than the original price. The buy orders before the burst of messages are possible from insiders, who are traders with previous information about the \pump occurrence (channel administrators or VIP members), as described by \cite{LaMorgia:2020}. Figure \ref{fig:signature-mr} describes a possible \texttt{\$BICO} market valuation trend guided by external events. The trading and pricing patterns are not aligned with short-lived \pump occurrences, but could be some long-term manipulation called \textit{Crowd-Pump}, which is a cooperative manipulation model that instigates the holding of the target asset within the goal of further valuation \cite{LaMorgia:2023}.

\emph{\textbf{Takeaway:}
(i) Telegram exhibits frequent anomalous attention bursts, but only a small subset translates into significant price and volume discontinuities, surfacing candidate \pump attempts at scale.
(ii) RDD–DiD validation separates coordinated \pump events from sustained, non-manipulative market reactions.}

\subsection{The Telegram Information Ecosystem}
\begin{figure}
    \centering
    \includegraphics[width=0.8\linewidth]{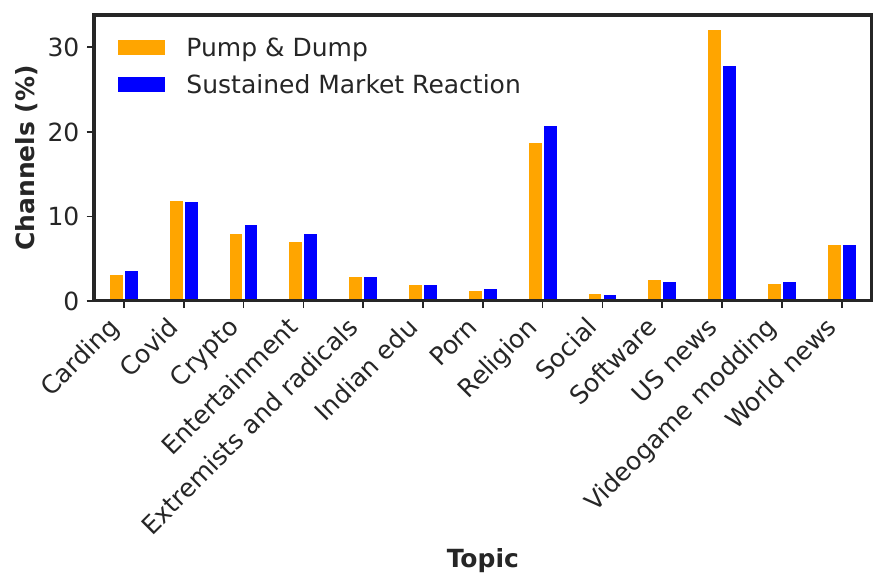}
    \caption{Channel Topics by Event Type (P\&D, SMR).}
    \label{fig:channel_dist}
\end{figure}

In terms of diversity of Telegram's ecosystem (\textbf{RQ3}), the P\&D cases accomplish 43,675 messages and 2,391 unique channels. The SMR group contains 7,210 messages and 2,081 channels. From this collection, 2,071 channels were present in both scenarios, indicating a high overlap in crypto-related content emissors. 410 channels were exclusive for emitting P\&D messages, while only 10 were present in SMR events. We used the provided channel's label defined by the dataset creators \cite{LaMorgia:2025} to explore the spreadness across different social spheres during crypto manipulation schemes or just commonplace discussions. As presented in Figure \ref{fig:channel_dist}, the majority of channels were from \textit{US News} and \textit{Religion} communities from both cases, with crypto groups being less than 10\% of the discussion spots. That highlights the \textit{Crypto's Everywhere phenomenon} observed by \citet{Paoletti:2025}, which is the acceptance of this theme across diverse social contexts, and not only being an echo chamber in social platforms. Other relevant topics are \textit{Covid} and \textit{Entertainement}.

\begin{figure}[t]
    \centering

    \begin{subfigure}{0.48\linewidth}
        \centering
        \includegraphics[width=\linewidth]{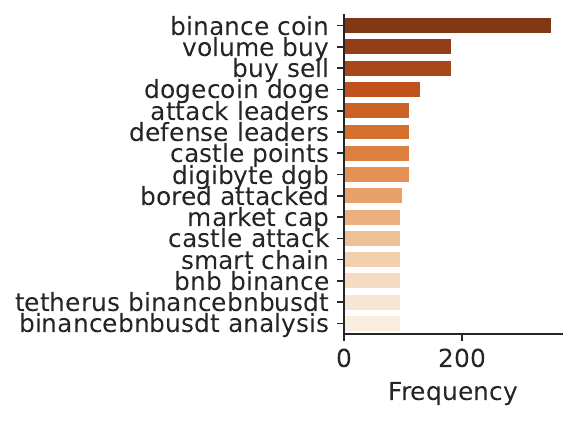}
        \caption{P\&D frequent bigrams.}
        \label{fig:pd-bigrams}
    \end{subfigure}
    \hfill
    \begin{subfigure}{0.48\linewidth}
        \centering
        \includegraphics[width=\linewidth]{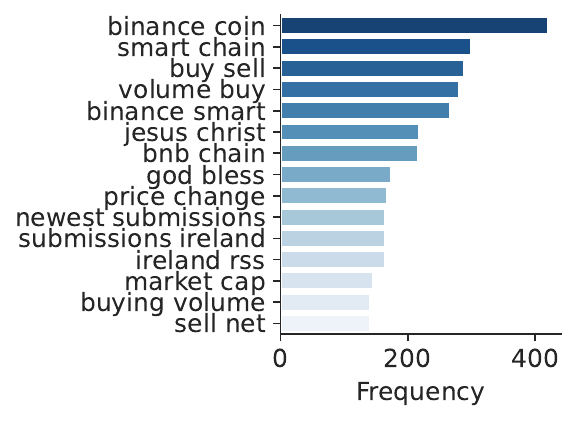}
        \caption{SMR frequent bigrams.}
        \label{fig:mr-bigrams}
    \end{subfigure}

    \caption{\rev{Most} frequent bigrams between P\&D and SMR cases.}
    \label{fig:frequent-bigrams}
\end{figure}

To characterize the distinctions of both cases, we explored the message's content, aiming to observe textual patterns that could differentiate P\&D instances from normal discussions, as only overlooking crypto channels seems not be sufficient. First, we observed the most frequent bigrams in each case, excluding classical stopwords, and also possible links included. Figure \ref{fig:frequent-bigrams} illustrates the most frequent bigrams for each case. As observed, both mainly contain the same dialect of incisive terms such as \textit{buy}, \textit{sell}, and \textit{buying}. Combinations such as \textit{Binance coin}, \textit{changing price}, \textit{smart chain} are commonly used terms in this environment as well. 
However, Figure~\ref{fig:pd-bigrams} shows that P\&D cases lack many non-crypto bigrams commonly observed in organic discussions (e.g., references to news feeds or public figures). Despite punctual mentions, the bigram frequency reveals the similarity in the text content for both cases, with the message volume as a disruptive differential. Examples of messages from P\&D and SMR can be found in Appendix D \ref{sec:app-llm}.

\begin{figure}
    \centering
    \includegraphics[width=0.6\linewidth]{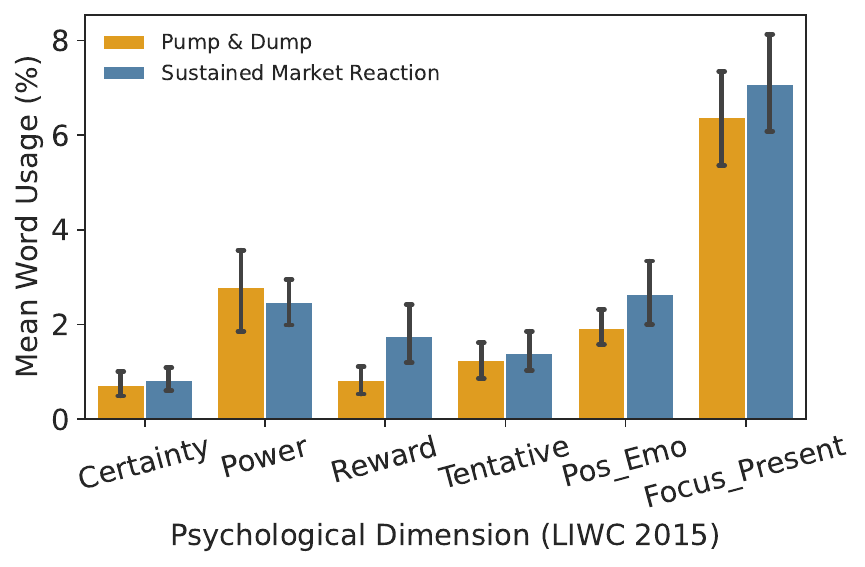}
    \caption{LIWC values for each category.}
    \label{fig:liwc-results}
\end{figure}

Beyond surface lexical patterns, we apply Linguistic Inquiry and Word Count (LIWC-2015)~\cite{Mohammadinodooshan:2025,Trinh:2025} to analyze the psycholinguistic properties of messages. LIWC maps words to validated linguistic, cognitive, and emotional categories; we focus on six dimensions commonly associated with persuasive, directive, and urgency-driven communication (Certainty, Power, Reward, Tentative, Pos\_Emo, and Focus\_Present), which prior work has linked to coordinated influence and financial persuasion. To reduce noise from repeated messages, we only consider unique messages per event. As shown in Figure~\ref{fig:liwc-results}, none of these dimensions significantly differ between P\&D and SMR messages.
Consistent with our bigram analyses, this shows that the linguistic surface of pump messages is largely indistinguishable from organic crypto discourse, revealing strong semantic mimicry and disputing the discriminative power of text-based analysis alone.

\paragraph{Temporal Latency and Precedence}
Since textual patterns alone fail to provide a clear demarcation between the two cohorts, \rev{the critical diagnostic dimension lies not in what is said but in when it is said relative to the market event it co-occurs with. Manipulation theory predicts a specific temporal ordering: in a \pump scheme, social signals serve as the trigger for price movement, meaning coordinated messaging must systematically precede the price peak. Conversely, in organic Sustained Market Reactions, social discussion is a response to market events, and messages should therefore concentrate after the price reaches its local maximum.}

\begin{figure}
    \centering
    \includegraphics[width=0.8\linewidth]{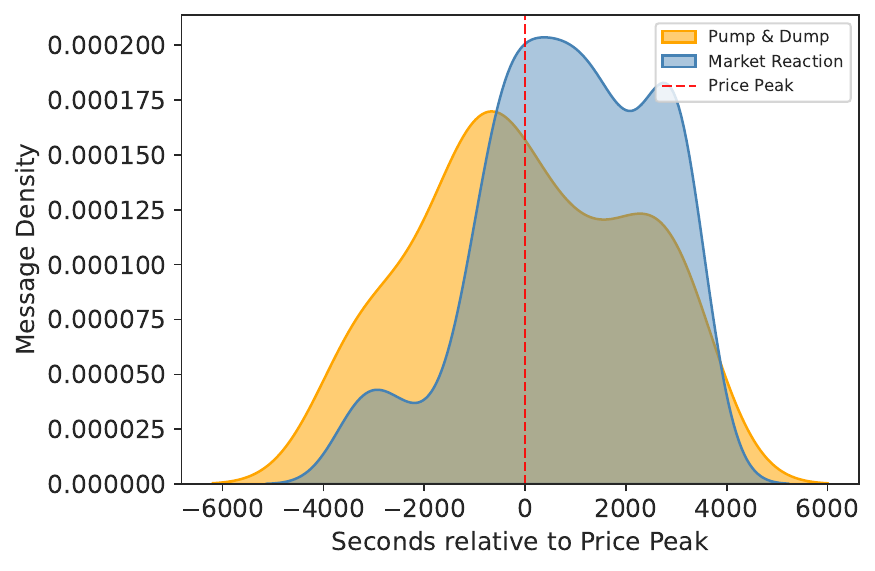}
    \caption{Latency Comparison between P\&D and SMR within $\pm1$ hour around the price peak.}
    \label{fig:latency-contrast}
\end{figure}

To operationalize this, we compute the latency of each message relative to the price peak of its associated event (the moment of maximum accumulated return within the observation window). Negative latency values indicate messages that were posted before the price peak; positive values indicate messages that followed it. Figure~\ref{fig:latency-contrast} illustrates the resulting distributions for \pump (N=47) and Sustained Market Reaction (N=73) events. The distributions diverge sharply in a manner consistent with the theoretical prediction. For \pump events, the median message latency is -$256$ seconds (-$4.3$ minutes), meaning that, at the midpoint of the distribution, coordinated social activity precedes the price peak by over four minutes. For Sustained Market Reactions, the median is +$868$ seconds (+$14.5$ minutes), indicating that social discussion concentrates after the price has already peaked. The difference between medians is $1,124$ seconds ($18.7$ minutes), a gap large enough to carry clear interpretive weight. A two-sample Kolmogorov-Smirnov test confirms that the two distributions are drawn from significantly different underlying processes (KS statistic = $0.272$, $p = 0.0006$), validating that the observed separation in temporal structure is not attributable to sampling noise. We report median rather than mean as the primary central tendency measure, given the heavy-tailed, non-normal character of both distributions.

As illustrated in Figure~\ref{fig:latency-contrast}, the P\&D density (orange) is left-skewed relative to the price peak, with substantial mass concentrated in the pre-peak window between approximately -$4,000$ and 0 seconds. The SMR density (blue) is bimodal, with a first mode just before the peak and a dominant second mode near +$3,000$ seconds, consistent with an initial reactive spike of attention at the price event followed by sustained community discussion as the new price level is processed and debated. Together, these temporal profiles constitute the primary behavioral signature distinguishing manipulation from organic market activity in our dataset. \pump events are characterized by social coordination that leads to price peaks; Sustained Market Reactions are characterized by social discussion that follows them. This precedence pattern -- confirmed across a dataset spanning 17,000 assets and 14,499 channels -- is the observable footprint of a manipulation phase that requires public amplification to secure exit liquidity before organic participants can react.

\emph{\textbf{Takeaway:}
(i) \pump activity spans many non-crypto Telegram communities, consistent with the \textit{Crypto’s Everywhere} phenomenon.
(ii) Linguistic content does not distinguish \pump events from organic market discussions.
(iii) Coordinated \pump events are instead characterized by extreme temporal synchronization of messages.}

\subsection{Economic Impact Analysis}

Finally, to measure the market dynamics of the detected cases, we quantify each validated event’s financial footprint in USD-equivalent terms using high-frequency minute-level price (open, high, low, close) and volume data.
For this analysis, we restrict the sample to deduplicated events for which
at least one USD-pegged or convertible trading pair was available across
the observation window. Volume figures are aggregated across all
(exchange, trading pair) combinations confirmed by the RDD--DiD step, using
time-weighted average price (TWAP) conversions for non-USD-quoted assets.
Price metrics are drawn from the most liquid USD-pegged pair per event.
Table~\ref{tab:event_stats} summarizes the resulting statistics.

\begin{table}[t]
\centering
\footnotesize
\setlength{\tabcolsep}{3.5pt}
\begin{tabular}{lcccccc}
\toprule
Cat. & Vol. (USD) & $ \Delta P^{+}$ (\%) & $\Delta P^{-}$ (\%) & $ \bar{t}_{\text{pump}}$ & $\bar{t}_{\text{dump}}$ & $N$ \\
\midrule
P\&D & \$234.8\,Mi & $+9.98$  & $-14.84$ & 14.6 & 32.1 & 47  \\
SMR  & \$828.8\,Mi & $+18.56$ & $-5.36$  & ---       & ---       & 73 \\
\bottomrule
\end{tabular}
\caption{Economic impact summary. $\Delta P^{+}$: average maximum price
increase from T\textsubscript{0} to the price peak. $\Delta P^{-}$:
average price loss for traders entering at the peak and holding to the
close (upper-bound loss estimate). $\bar{t}_{\text{pump}}$ and
$\bar{t}_{\text{dump}}$: mean duration of the pump and dump phases (minutes),
respectively. All values pegged to USD.}
\label{tab:event_stats}
\end{table}

The volume figures reveal a pattern consistent with the structural
differences identified in the temporal analysis. SMR events are associated
with substantially higher aggregate volume (\$828.8 million) than P\&D
events (\$234.8 million). This disparity, however, reflects the nature of
the assets involved rather than the intensity of the anomaly: organic market
reactions tend to cluster around higher-liquidity coins with established
trading depth, whereas P\&D schemes systematically target low-capitalization
assets where coordinated buy pressure can move prices with far smaller
capital.

The most diagnostically significant metrics are the asymmetric price
dynamics. P\&D events exhibit a mean maximum price increase of 9.98\%
from $T_0$ to the price peak, a gain that is entirely reversed and then exceeded by a mean maximum drawdown of $-14.84\%$ for participants who entered at the peak. This sign reversal constitutes the defining economic harm of the scheme: the coordination mechanism artificially sustains a price level long enough to attract external
buyers, who then absorb the losses when early participants exit. In contrast, SMR events show a higher average price gain (18.56\%) that
decays into a modest drawdown of only $-5.36\%$, consistent with a market processing new information and converging to a new equilibrium rather than collapsing back below the pre-event baseline.

The temporal structure of P\&D events further corroborates this
interpretation. The mean pump phase ($14.6$~minutes) is less than half the
mean dump phase (32.1~minutes), producing a Pump-to-Dump ratio below 0.5.
This asymmetry captures the mechanics of the scheme: the pump is executed
rapidly through coordinated social signals, maximizing the speed of price
inflation before organic participants can react, while the dump unfolds
more gradually as early sellers unwind positions against declining
liquidity. The extended dump phase is also consistent with the lingering
tail of post-peak social activity identified in the temporal latency
analysis, where the positive skewness of P\&D message distributions
reflects the residual shilling and panic-selling that characterize the
scheme's secondary phase.

Together, these metrics provide economic validation of our framework's
categorical separation. P\&D events are not simply more volatile than SMR
events, they exhibit a structurally distinct pattern of rapid,
artificial price inflation followed by a collapse that systematically
transfers value from late buyers to early sellers, precisely as predicted
by market microstructure theory.

\emph{\textbf{Takeaway:} \pump events target low-liquidity assets and generate rapid, artificial price inflation (averaging a 9.98\%
coordinated gain) followed by a collapse that exposes peak buyers to losses exceeding 14.84\%. The pump phase completes in under 15 minutes on average, while the dump unfolds over twice as long, consistent with sequential exit by coordinated early sellers against diminishing organic liquidity.}

\section{Conclusion and Future Works}

In this study, we presented a multi-layered framework for identifying and characterizing market anomalies mediated by social media, specifically focusing on the Telegram ecosystem. By integrating NLP-based semantic filtering (RoBERTa), statistical burst detection (CA-CFAR), and rigorous econometric validation (RDD-DiD), we identified 47 candidate \pump events
and 73 Sustained Market Reactions from a dataset spanning 17,000 assets.  Our results reveal a critical finding: while manipulators employ semantic mimicry that renders their messages linguistically indistinguishable from organic discourse, they are betrayed by a distinct temporal signature, coordinated social activity that systematically \emph{precedes} price peaks by a median of 4.3 minutes, whereas organic market reactions follow price peaks by a median of 14.5 minutes. This directional asymmetry, confirmed by a Kolmogorov-Smirnov test ($p = 0.0006$), provides content-independent evidence consistent with social signs leading price changes in manipulation events, and price changes leading social discussion in organic events. Economically, \pump events target low-liquidity assets and generate rapid artificial
inflation -- averaging a 9.98\% coordinated gain from $T_0$ to the price peak -- followed by a collapse that exposes peak buyers to losses exceeding 14.84\%, nearly three times the 5.36\% drawdown observed in organic Sustained Market Reactions. This asymmetric harm profile, where the pump benefits early coordinated sellers at the direct expense of late buyers, points to a systemic risk to market fairness in decentralized cryptocurrency markets.

This research bridges a gap in the literature by extending the detection of \pump schemes beyond specialized echo chambers into a broader social ecosystem. By implementing a classification layer that resolves semantic polysemy in cryptocurrency identifiers, our work accounts for the \textit{Crypto's Everywhere} phenomenon, where manipulative signals diffuse
across diverse social spheres, from crypto channels to news and entertainment communities. Unlike prior studies restricted to known predatory groups, our approach combines econometric validation with temporal profiling to identify coordinated anomalies amidst general-purpose
social discourse, while also distinguishing deliberate \pump orchestrations
from sustained market reactions to emerging trends.

As for future work, we intend to explore coordination patterns more deeply, analyzing the network topology of cross-channel synchronization and the role of automated accounts in amplifying signals. \rev{We also plan to extend our analysis to public Telegram groups beyond the channels considered in this study, broadening the scope of observable coordination signals.}
A major objective is the construction of a comprehensive, multimodal ground-truth dataset for \pump occurrences, fusing verified social media commands with high-frequency order book data to provide the research community with a benchmark for training more resilient detection models.

\subsection{Limitations}
While our multi-modal framework demonstrates robust detection capabilities, we acknowledge certain scope boundaries. \rev{First, our primary data source consists of Telegram public channels, which utilize a broadcast communication model. While it is highly likely that the initial orchestration of these schemes occurs within closed private groups/channels, the intrinsic mechanics of a pump-and-dump scheme necessitate public exposure to secure exit liquidity. Therefore, while planning may be private, the execution may generate a massive footprint in public channels. The CA-CFAR algorithm is particularly well-suited to capture these high-intensity signal injections.}

Second, the RDD-DiD framework is inherently reactive, identifying anomalies after the initial social signal. However, our temporal analysis demonstrates that manipulative bursts precede price movements by mere seconds, suggesting that even reactive detection can inform near-real-time monitoring systems. The reliance on stablecoins and Bitcoin control groups to filter systemic noise is robust under normal conditions; during extreme market-wide volatility, correlation shifts may introduce bias, which we mitigate by excluding dates with documented systemic shocks. Finally, the semantic mimicry identified in this work poses an ongoing challenge. As manipulators evolve, they may adjust their temporal patterns to evade detection. This underscores the need for continuous model adaptation and motivates our future work on multimodal ground-truth datasets that can capture evolving manipulation strategies.

\bibliography{aaai2026}

\section{Paper Checklist}

\appendix

\begin{enumerate}

\item For most authors...
\begin{enumerate}
    \item  Would answering this research question advance science without violating social contracts, such as violating privacy norms, perpetuating unfair profiling, exacerbating the socio-economic divide, or implying disrespect to societies or cultures?
    \answerYes{Yes}
  \item Do your main claims in the abstract and introduction accurately reflect the paper's contributions and scope?
    \answerYes{Yes}
   \item Do you clarify how the proposed methodological approach is appropriate for the claims made? 
    \answerYes{Yes}
   \item Do you clarify what are possible artifacts in the data used, given population-specific distributions?
    \answerNA{NA}
  \item Did you describe the limitations of your work?
    \answerYes{Yes, see section Limitation.}
  \item Did you discuss any potential negative societal impacts of your work?
    \answerNA{NA}
      \item Did you discuss any potential misuse of your work?
    \answerNA{NA}
    \item Did you describe steps taken to prevent or mitigate potential negative outcomes of the research, such as data and model documentation, data anonymization, responsible release, access control, and the reproducibility of findings?
    \answerNA{NA}
  \item Have you read the ethics review guidelines and ensured that your paper conforms to them?
    \answerYes{Yes}
\end{enumerate}

\item Additionally, if your study involves hypotheses testing...
\begin{enumerate}
  \item Did you clearly state the assumptions underlying all theoretical results?
    \answerYes{Yes}
  \item Have you provided justifications for all theoretical results?
    \answerYes{Yes}
  \item Did you discuss competing hypotheses or theories that might challenge or complement your theoretical results?
    \answerNA{NA}
  \item Have you considered alternative mechanisms or explanations that might account for the same outcomes observed in your study?
    \answerYes{Yes. We account for alternative explanations including market-wide shocks and organic information diffusion by using RDD combined with DiD and stablecoin controls, and we interpret our findings as correlational rather than strictly causal.}
  \item Did you address potential biases or limitations in your theoretical framework?
    \answerYes{Yes}
  \item Have you related your theoretical results to the existing literature in social science?
    \answerNA{NA}
  \item Did you discuss the implications of your theoretical results for policy, practice, or further research in the social science domain?
    \answerYes{Yes}
\end{enumerate}

\item Additionally, if you are including theoretical proofs...
\begin{enumerate}
  \item Did you state the full set of assumptions of all theoretical results?
    \answerNA{NA}
	\item Did you include complete proofs of all theoretical results?
    \answerNA{NA}
\end{enumerate}

\item Additionally, if you ran machine learning experiments...
\begin{enumerate}
  \item Did you include the code, data, and instructions needed to reproduce the main experimental results (either in the supplemental material or as a URL)?
    \answerYes{Yes. We will release a public GitHub repository containing the full experimental code and detailed instructions to reproduce the main results, along with the cryptocurrency dictionary collected in this study. The Telegram data used in our analysis are already publicly available, and the repository will document how to access and preprocess them.}
  \item Did you specify all the training details (e.g., data splits, hyperparameters, how they were chosen)?
    \answerYes{Yes}
     \item Did you report error bars (e.g., with respect to the random seed after running experiments multiple times)?
    \answerNA{NA}
	\item Did you include the total amount of compute and the type of resources used (e.g., type of GPUs, internal cluster, or cloud provider)?
    \answerNo{No}
     \item Do you justify how the proposed evaluation is sufficient and appropriate to the claims made? 
    \answerYes{Yes}
     \item Do you discuss what is ``the cost`` of misclassification and fault (in)tolerance?
    \answerNA{NA}
  
\end{enumerate}

\item Additionally, if you are using existing assets (e.g., code, data, models) or curating/releasing new assets, \textbf{without compromising anonymity}...
\begin{enumerate}
  \item If your work uses existing assets, did you cite the creators?
    \answerYes{Yes}
  \item Did you mention the license of the assets?
    \answerNA{NA}
  \item Did you include any new assets in the supplemental material or as a URL?
    \answerNo{No}
  \item Did you discuss whether and how consent was obtained from people whose data you're using/curating?
    \answerYes{Yes. This study relies exclusively on an existing public dataset from the literature, consisting of messages collected from Telegram public channels. The dataset was originally gathered via the Telegram API and includes only content that is openly accessible by design, where users have no reasonable expectation of privacy. Following established ethical guidelines for computational social science research, individual consent was not sought, as the data are public, non-interactive, and analyzed in aggregate form.}

  \item Did you discuss whether the data you are using/curating contains personally identifiable information or offensive content?
    \answerYes{Yes. The data consist of public Telegram channel content and may include user-generated material, but we do not curate the dataset nor analyze personally identifiable information, focusing only on aggregated signals.}
\item If you are curating or releasing new datasets, did you discuss how you intend to make your datasets FAIR (see \citet{fair})?
\answerNA{NA}
\item If you are curating or releasing new datasets, did you create a Datasheet for the Dataset (see \citet{gebru2021datasheets})? 
\answerNA{NA}
\end{enumerate}

\item Additionally, if you used crowdsourcing or conducted research with human subjects, \textbf{without compromising anonymity}...
\begin{enumerate}
  \item Did you include the full text of instructions given to participants and screenshots?
    \answerNA{NA}
  \item Did you describe any potential participant risks, with mentions of Institutional Review Board (IRB) approvals?
    \answerNA{NA}
  \item Did you include the estimated hourly wage paid to participants and the total amount spent on participant compensation?
    \answerNA{NA}
   \item Did you discuss how data is stored, shared, and deidentified?
   \answerNA{NA}
\end{enumerate}

\end{enumerate}

\section{Ethical Considerations}

This study relies exclusively on observational data and does not involve any interaction or intervention with human subjects. All Telegram data analyzed originate from publicly accessible Telegram channels and were collected from an existing dataset released by prior work. The authors did not access private groups, bypass platform restrictions, or engage with users in any form.

We adopt a strict de-identification approach and do not attempt to identify, track, or profile individual users, administrators, or coordinated actors. All analyses are performed at the aggregate level (messages, channels, and events), and no personally identifiable information is collected or inferred.

Market data (prices and transaction volumes) are obtained from public sources through official APIs and collected in accordance with their documented usage policies.

While we identify patterns consistent with potential pump-and-dump activity, our findings are descriptive and focused on characterizing market anomalies. They do not constitute legal evidence of intent or wrongdoing by specific individuals, but rather serve to highlight the systemic risks that coordinated social media activity poses to the fairness of decentralized financial markets.

\section{Appendix A - Language Models} \label{sec:app-llm}
This section describes the Hyperparameters used in the evaluated models, the prompt templates used during ICL inferences, and the generalization/robustness validation experiment

\paragraph{Hyperparameters:} To identify the optimal configuration for our classification models, we conducted an extensive hyperparameter search (Grid-Search) using a Bayesian optimization approach (Optuna). The search space was designed to balance model stability with domain adaptation, exploring the following ranges: learning\ rate $\in \{1\times10^{-5},\ 2\times10^{-5},\ 3\times10^{-5},\ 5\times10^{-5}\}$, training epochs $\in$ \{2,3,4\}, weight decay $\in$ \{0.0,0.01,0.1\}, and dropout rate $\in$\{0.1,0.2,0.3\}. All trials were evaluated using the validation split of our ground-truth dataset.

The \texttt{RoBERTa-base} model, which was subsequently selected as our primary classifier, achieved its peak performance (Objective Score: 0.9429) with the following optimized hyperparameters: a learning rate of 3×10-5, 4 training epochs, a weight decay of 0.01, and a dropout rate of 0.1. These settings ensured the best trade-off between capturing specialized cryptocurrency jargon and maintaining generalisation across different Telegram channel styles, effectively mitigating the risk of overfitting during the fine-tuning process.

\begin{table}[t]
\centering
\small
\setlength{\tabcolsep}{3pt}
\renewcommand{\arraystretch}{1.05}

\begin{tabular}{p{1.8cm}cccc}
\toprule
\textbf{Model} & \textbf{ZS} & \textbf{ICL Z} & \textbf{ICL F} & \textbf{FT} \\
\midrule
gemma-7b
& $\approx$53k
& $\approx$65k
& $\approx$116k
& 69.47 (554.26$\pm$44.03) \\
llama3.1-8B
& $\approx$55k
& $\approx$67k
& $\approx$125k
& 63.63 (503.36$\pm$34.11) \\
qwen2.5-8B
& $\approx$47k
& $\approx$56k
& $\approx$100k
& 87.79 (831.27$\pm$35.49) \\
roberta
& --
& --
& --
& 4.78 (89.14$\pm$8.40) \\
\bottomrule
\end{tabular}

\caption{\footnotesize
Average inference time (ms) per message across strategies. 
FT column includes fine-tuning time (mean$\pm$95\% CI).}
\label{tab:time_comparison_update}
\end{table}

Beyond classification effectiveness, we also assess the computational implications of the selected hyperparameter configuration. Table~\ref{tab:time_comparison_update} summarizes the average inference time per message across models and training strategies, contextualizing the scalability of the proposed approach.

\paragraph{Prompt Templates:} We here present the prompts used for Zero-Shot Classification, Zero-Shot In-Context Learning (ICL), and Few-Shot In-Context Learning (ICL). The Zero-Shot Classification prompt is nearly identical to the Zero-Shot ICL prompt, except that it does not include the contextual guideline describing the criteria for a message to be considered crypto-related.

\begin{msgbox}{Zero-Shot In-Context Learning,breakable}
You are an expert in analyzing Telegram cryptocurrency communication.\newline
\newline
Classify the message strictly as:\newline
1 = Crypto-related\newline
0 = Not crypto-related\newline
\newline
A message is crypto-related if it mentions:\newline
- Cryptocurrencies (Bitcoin, Ethereum, altcoins, tokens)\newline
- NFTs, blockchain, Web3, DeFi, exchanges, wallets, mining, staking, airdrops\newline
- Crypto trading, investing, tokens, presales, jargons or memecoins\newline
\newline
Even a brief or indirect mention is sufficient.\newline
\newline
Message:\newline
\{message\}\newline
\newline
Answer only with 0 or 1:
\end{msgbox}

\begin{msgbox}{Few-Shot In-Context Learning (5examples),breakable}
You are an expert in cryptocurrency communication analysis.\newline
\newline
Classify the message strictly as:\newline
1 = Crypto-related\newline
0 = Not crypto-related\newline
\newline
Here are some examples:\newline
Message: \{message\}\newline
Label: \{label\}\newline
\newline Now classify the next message:\newline
\newline
Message: \{message\}\newline
\newline
Answer only with 0 or 1:
\end{msgbox}

\begin{figure*}[t]
    \centering
    \includegraphics[width=0.8\linewidth]{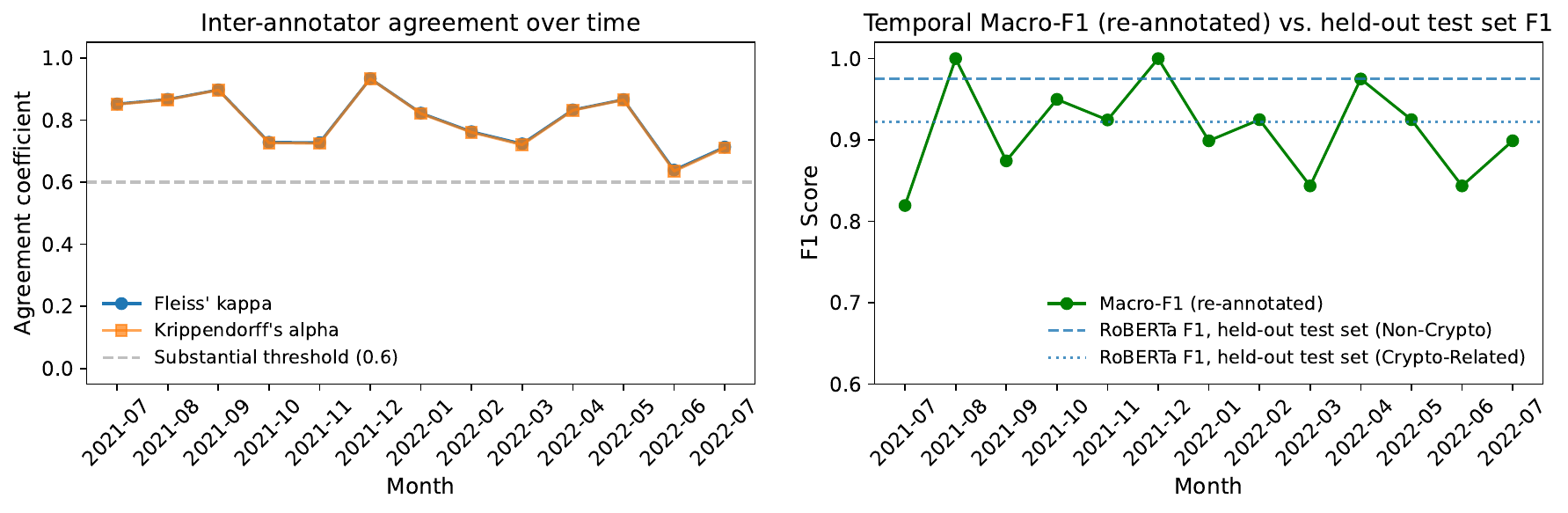}
    \caption{Annotation agreement evaluation (left) and generalization experiment (right)}
    \label{fig:annotation-exp}
\end{figure*}

\rev{\paragraph{Robustness and Generalization:} 
To further validate the temporal robustness and generalization capabilities of our fine-tuned RoBERTa model, we conducted a longitudinal annotation experiment. We extracted a temporally stratified sample of 520 messages spanning the 13 months of our analysis dataset (40 messages per month, equally balanced between predicted positive and negative classes). Three independent annotators, strictly blinded to the model's predictions, manually classified these samples following our original labeling codebook. To evaluate the reliability of this new ground truth, we computed both Fleiss' Kappa and Krippendorff's Alpha. As shown in Figure \ref{fig:annotation-exp} (left), both metrics yielded nearly identical values and consistently exceeded the substantial agreement threshold ($0.6$) across all months. This minimal divergence and high sustained value confirm a strong, reliable consensus among the human annotators over time. Furthermore, when evaluated on this longitudinally re-annotated data, the model demonstrated remarkable temporal stability. As illustrated in Figure \ref{fig:annotation-exp} (right), the monthly Macro-F1 scores remain consistent across the entire observation period for both the crypto-related and non-crypto classes, remaining consistent with the performance observed on the held-out test set. These results empirically demonstrate that our semantic filter does not suffer from temporal degradation, ensuring robust and generalized detection across the entire observation period of this study.}

\section{Appendix B - CA-CFar Sensibility Analysis} \label{sec:app-cfar}
To calibrate the CA-CFAR algorithm, we performed a sensitivity analysis on two critical parameters: Guard Cells and Training Cells. To establish a reference for this tuning, we utilized a subset of verified \pump (P\&D) events annotated by \citet{LaMorgia:2020}. Although this external dataset lacks the original message content, we successfully cross-referenced two major channels—\textit{Whales Crypto Guide} (WCG) and \textit{Coach} (formerly \textit{Crypto coin B}) -- to our repository. After filtering for temporal consistency, we isolated 16 confirmed P\&D events from the WCG channel (2018–2020) to serve as a validation ground-truth. While these events precede our primary analysis window (2021–2022), they provide the necessary structural patterns of social bursts required for parameter optimization.

We tested a range of values for both parameters: \{1,2,3,5,7,15,30\} days. The Guard Cells define the period immediately surrounding the analysis day that must be ignored to prevent the signal burst from corrupting the baseline, while the Training Cells establish the historical window used to calculate the noise floor. Our analysis revealed that a configuration of 1 Guard Cell and 5 Training Cells achieved the highest sensitivity, successfully detecting 15 out of 16 ground-truth events.

The preference for these specific values is driven by the intrinsic nature of social media manipulation. We observed that increasing the Guard Cell window led to a significant drop in detection rates; this is likely because the highly concentrated, short-lived nature of social bursts is absorbed or ``smoothed over'' by excessively long exclusion zones, which fail to isolate the localized discontinuity of a pump signal. Similarly, a 5-day Training Cell window provided the optimal balance for the noise floor, long enough to establish a stable baseline of organic discussion, yet short enough to remain responsive to the rapid shifts in cryptocurrency social trends. This configuration ensures that the algorithm remains highly sensitive to sudden, synchronized bursts while filtering out systemic background noise.

\section{Appendix C - Causality Methods Evaluation} \label{sec:app-causality}
While this study primarily relies on the synergy between Regression Discontinuity Design (RDD) and Difference-in-Differences (DiD), we also explored Interrupted Time Series (ITS) and Convergent Cross Mapping (CCM). This appendix details their mechanics and justifies why RDD and DiD remain the superior choices for detecting high-frequency market manipulations. Finally, we demonstrate the threshold sensitivity during \pump and Sustained Market Reaction classifications during the RDD + DiD framework.
\subsubsection{Interrupted Time Series - ITS}
ITS is a quasi-experimental design that models a series of observations over time, interrupted by a specific event. Unlike RDD, which focuses on the instantaneous ``jump'' at the threshold, ITS is designed to evaluate the slope change and long-term trends before and after an intervention. It is ideal for capturing the ``persistence'' of a shock. It excels at identifying if a market reaction has ``long-memory'' properties or if the intervention caused a permanent shift in the baseline trading volume. The primary weakness of ITS in the \pump context is its reliance on a stable pre-intervention trend. Cryptocurrency markets are notoriously volatile; a 120-minute window often contains secondary shocks that contaminate the slope estimation. While ITS provides a good view of Volume Decay, it lacks the ``local precision'' of RDD, which is less sensitive to distant noise within the time series.

\subsubsection{Convergent Cross Mapping (CCM)}
CCM is a non-linear state-space method used to detect causal coupling in complex systems. It reconstructs the shadow manifolds (attractors) of two variables (e.g., Telegram message volume and Price returns) to see if information from one is embedded in the other. Unlike Granger Causality, CCM can detect ``pre-event leakage'' or ``insider trading'' where social media activity forces price movements in a non-linear, coupled fashion before the official $T_0$. In the case of this study, it presented the following challenges and limitations: (i) Model Fitting and Stability: During our implementation, CCM presented significant fitting problems. Financial time series are often non-stationary and contain high levels of stochastic noise, which ``shatters'' the manifold reconstruction, leading to poor convergence. (ii) Scalability and Retraining: CCM is inherently asset-specific. A manifold reconstruction model must be individually trained and tuned for every single coin in the dataset. Given the thousands of anomalous events detected across hundreds of different altcoins, a per-coin CCM approach is computationally prohibitive and less viable for a generalized detection framework compared to the RDD + DiD strategy.

\begin{figure}[htbp]
    \centering
    \begin{subfigure}[b]{0.6\linewidth}
        \centering
        \includegraphics[width=\linewidth]{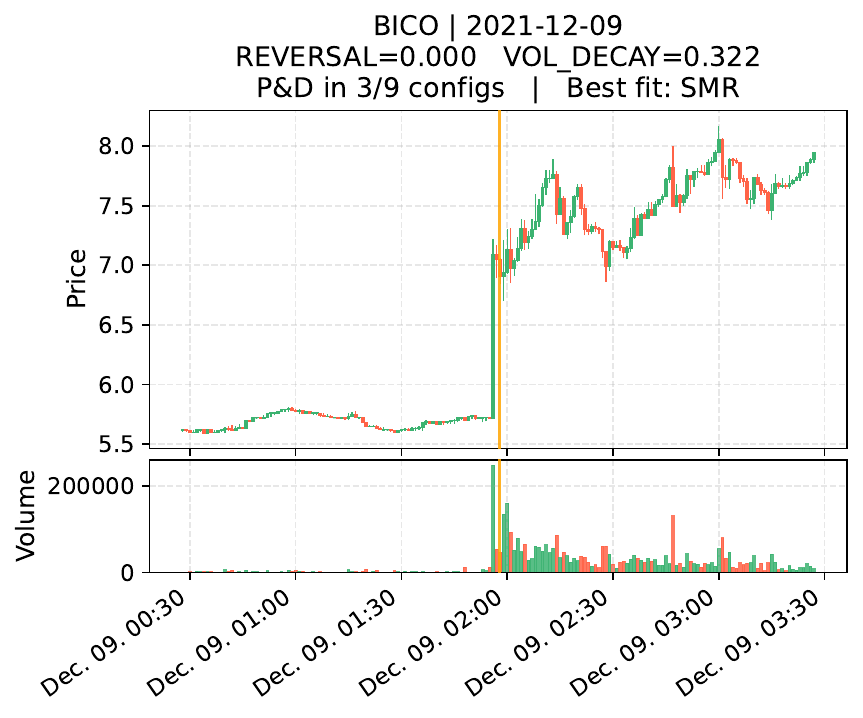}
        \caption{BICO (2021-12-09)}
        \label{fig:boundary_bico}
    \end{subfigure}
    
    
    \begin{subfigure}[b]{0.6\linewidth}
        \centering
        \includegraphics[width=\linewidth]{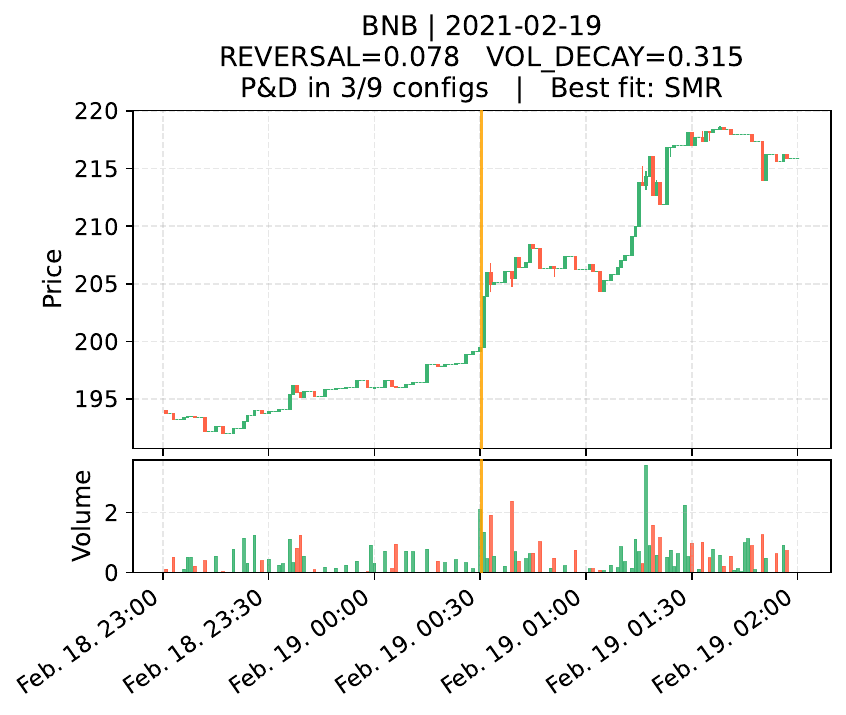}
        \caption{BNB (2021-02-19)}
        \label{fig:boundary_bnb1}
    \end{subfigure}


    \begin{subfigure}[b]{0.6\linewidth}
        \centering
        \includegraphics[width=\linewidth]{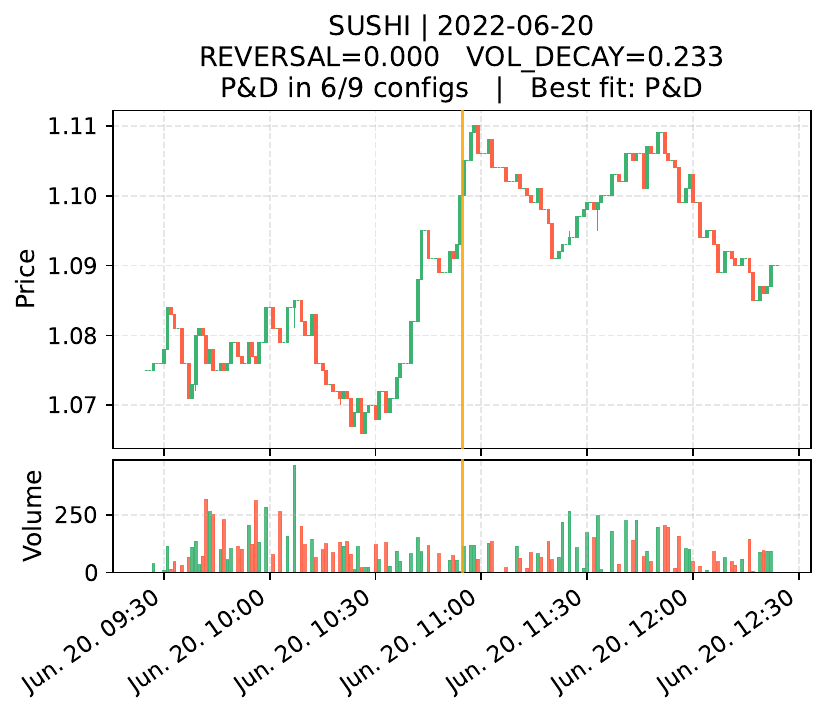}
        \caption{SUSHI (2022-06-20)}
        \label{fig:boundary_bnb2}
    \end{subfigure}
    
     
    \begin{subfigure}[b]{0.6\linewidth}
        \centering
        \includegraphics[width=\linewidth]{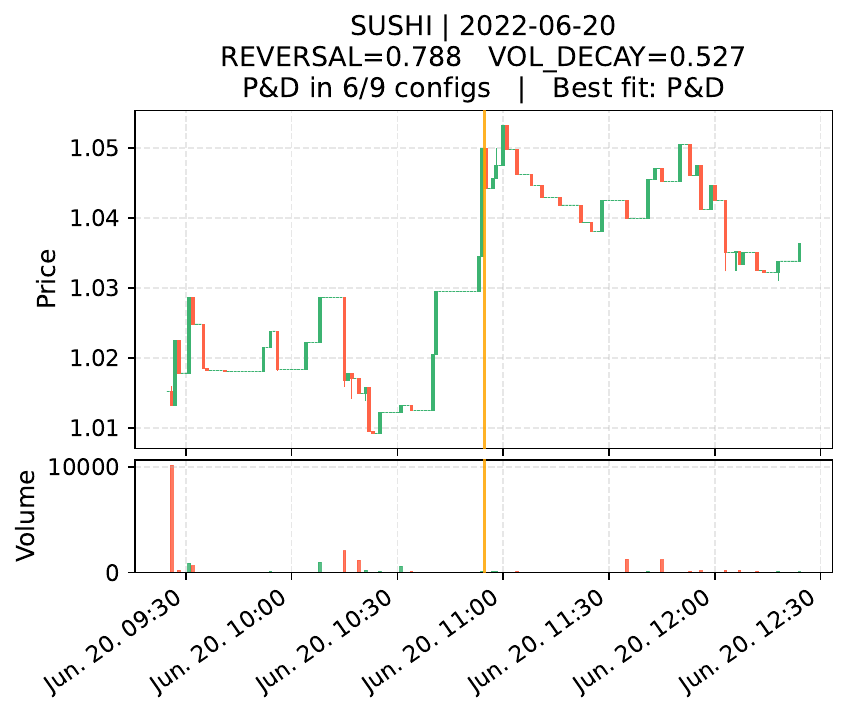}
        \caption{SUSHI (2022-06-20) in another exchange}
        \label{fig:boundary_clv}
    \end{subfigure}

    \caption{
    Manual inspection of borderline cases. \texttt{\$BICO} and \texttt{\$BNB} are SMR events misclassified as \pump under some threshold settings, due to transient corrections or temporary volume drops. The \texttt{\$SUSHI} panels show staged P\&D patterns across two exchanges, where the dump phase appears as discrete bursts rather than a single decline.}
    
    \label{fig:appendix_boundary_cases}
\end{figure}

\rev{\subsubsection{Thresold Sensitiviy Analysis}
To empirically validate the robustness of our classification thresholds, we conducted a comprehensive sensitivity analysis. We evaluated nine distinct parameter configurations by varying the Reversal Ratio across $0.6, 0.7$, and $0.8$, and the Volume Decay across $0.2, 0.3$, and $0.4$. The categorization of events remained highly stable across the majority of these combinations, demonstrating that our pipeline is not overly sensitive to minor parameter adjustments. However, we identified 18 (15\%) borderline cases where the event classification (\pump vs. Sustained Market Reaction) fluctuated depending on the specific configuration applied. To resolve these discrepancies, we conducted a rigorous manual qualitative inspection of the high-frequency financial time series for these 18 boundary events. Our analysis revealed that deviating from the baseline $0.7$ and $0.3$ configuration frequently introduced false positives into both categories. As presented by Table \ref{tab:boundary-events-sensibility}, for instance, alternative values either misclassified partial-dump manipulations as organic price discovery or flagged legitimate sustained reactions exhibiting natural trading lulls as artificial dumps. Consequently, we determined that retaining the $0.7$/$0.3$ configuration provides the most accurate and theoretically sound separation. Selected examples of these manually inspected boundary cases -- such as BICO, BNB, and SUSHI (in two different exchanges) -- are presented in Figure \ref{fig:appendix_boundary_cases} to illustrate how true sustained reactions can be prematurely flagged as manipulation under suboptimal threshold configurations, and how \pump can be missclassified as sustained market reactions even with abrupt pre-burst buy orders followed by different \textit{pump} stages, which were sensitive to the Volume Decay threshold.}

\begin{table}[t]
\centering

\resizebox{\columnwidth}{!}{%
\begin{tabular}{c|cc|cc|cc}
\toprule
\textbf{Vol.} 
  & \multicolumn{2}{c|}{\textbf{0.2}} 
  & \multicolumn{2}{c|}{\textbf{0.3}} 
  & \multicolumn{2}{c}{\textbf{0.4}} \\

\cmidrule(lr){2-3}\cmidrule(lr){4-5}\cmidrule(lr){6-7}
\textbf{Rev.}  & \textit{P\&D} & \textit{SMR} 
  & \textit{P\&D} & \textit{SMR} 
  & \textit{P\&D} & \textit{SMR} \\
\textbf{0.6} &  3 & 15 & 11 &  7 & 18 &  0 \\
\textbf{0.7} &  0 & 18 &  \textbf{8} & \textbf{10} & 15 &  3 \\
\textbf{0.8} &  0 & 18 & 13 &  5 & 15 &  3 \\
\bottomrule
\end{tabular}
}
\caption{Sensitivity of boundary event classification ($n=18$) across threshold configurations.}
\label{tab:boundary-events-sensibility}
\end{table}


\section{Appendix D - Messages Examples}
This section includes examples of messages during abrupt events (\pump and Sustained Market Reactions).
\subsection{Pump \& Dump Validated Samples}

\begin{msgbox}{P\&D Example 1: Coordination and Historical Returns}
\#RVN \#1 (BTC) [Bittrex] \\
\textbf{15 min burst detected} \\
Vol24h: 12.55 (+7.38 BTC) 142.8\% \\
Price Change: +14.81\% \\
\textit{Previous events:} 10 Aug +68.92\% (15 min); 06 Aug +59.45\% (15 min). \\
RVN - Ravencoin (CMC \#83)
\end{msgbox}

\begin{msgbox}{P\&D Example 2: Signal Accuracy and ROE Shilling}
4 XBT signals in last 48 hours with 100\% Accuracy. \\
\#XBT 1: 140\% ROE | \#XBT 3: 214\% ROE \\
\#BAT: 35\% Profit | \#CHZ: 151\% Profit \\
Exchanges: Binance.
\end{msgbox}

\begin{msgbox}{P\&D Example 3: Influencer-Led Urgency (FOMO)}
DOGE COIN HEADING SKYROCKET. Elon Musk Promoted this Early Morning. Invest and Earn a Big stack. Move your btc to DOGE. Buy and Sell using [URL].
\end{msgbox}

\subsection{Sustained Market Reaction Samples}

\begin{msgbox}{Market Reaction 1: Infrastructure and Rebranding}
First it was Binance chain, then Binance smart chain and now its \textbf{BNB} (Build \& Build) \textbf{chain}. Official Support Link: [URL]
\end{msgbox}

\begin{msgbox}{Market Reaction 2: Whale Activity and Liquidity Flows}
12,499,999 \#XLM (2,722,654 USD) transferred from unknown wallet to \#Coinbase. [Details on Whale Alert].
\end{msgbox}

\begin{msgbox}{Market Reaction 3: Organic Community Speculation}
Maybe if we are lucky we can convince Elon Musk to send all of us to another universe on his rocket with all of the Dogecoin so that we can escape this insanity.
\end{msgbox}

\end{document}